%% file: main.tex
\documentclass[journal]{IEEEtran}
\input{sections/preamble}
\graphicspath{./figs/}
\begin{document}
\bibliographystyle{unsrt}
\input{sections/title}
\input{sections/introduction}
\input{sections/preliminary}

\input{sections/motivation}

\input{sections/simulations}

\input{sections/conclusions}
\input{sections/ackowledgment}
\bibliography{main}
\end{document}

%% file: sections/preamble.tex
\IEEEoverridecommandlockouts
\usepackage{cite}
\usepackage{subcaption}
\usepackage{amsmath,amsfonts,amsthm,bm,amssymb} %
\usepackage[table,xcdraw]{xcolor}
\usepackage{algorithm}
\usepackage{algpseudocode} 
\usepackage{graphicx}
\usepackage{graphics}
\usepackage{epsfig}

\usepackage{footnote}
\usepackage{textcomp}
\PassOptionsToPackage{no-math}{fontspec}
\usepackage{tikz}  
\usepackage{makecell}
\usepackage{bm}
\usetikzlibrary{calc} 
\usetikzlibrary{arrows,shapes,chains} 
\usetikzlibrary{positioning, shapes.geometric}
\usepackage{multirow} 
\usepackage{verbatim}
\usepackage{enumitem}
\def\BibTeX{{\rm B\kern-.05em{\sc i\kern-.025em b}\kern-.08em
		T\kern-.1667em\lower.7ex\hbox{E}\kern-.125emX}}
	
\usepackage{verbatim}
\usepackage{pgfplots}
\usepackage{hyperref}
\pgfplotsset{width=10cm,compat=1.9}
\pgfplotsset{
	every axis legend/.style={
		cells={anchor=center},
		inner xsep=3pt,inner ysep=2pt,
		nodes={inner sep=2pt,text depth=0.1em},
		anchor=north east,
		shape=rectangle,
		fill=white,draw=green,
		font=\footnotesize
	},
}

\usepackage{verbatim}
\usepackage{booktabs}
\usepackage{fancyhdr}
\usepackage{svg}
\normalsize
	

%% file: sections/title.tex
\title{Deep learning based enhancement of ordered statistics decoding of short LDPC codes\\
}

\author{Guangwen Li, Xiao Yu

\thanks{G.Li is with the College of Information \& Electronics, Shandong Technology and Business University, Yantai, China e-mail: lgw.frank@gmail.com}
\thanks{X.Yu is with the Department of Physical Sports, Binzhou Medical University, Yantai, China e-mail: YuXiao@bzmu.edu.cn}
}

\maketitle
\begin{abstract}
In the search for highly efficient decoders for short LDPC codes approaching maximum likelihood performance, a relayed decoding strategy, specifically activating the ordered statistics decoding process upon failure of a neural min-sum decoder, is enhanced by instilling three innovations. Firstly, soft information gathered at each step of the neural min-sum decoder is leveraged to forge a new reliability measure using a convolutional neural network. This measure aids in constructing the most reliable basis of ordered statistics decoding, bolstering the decoding process by excluding error-prone bits or concentrating them in a smaller area. Secondly, an adaptive ordered statistics decoding process is introduced, guided by a derived decoding path comprising prioritized blocks, each containing distinct test error patterns. The priority of these blocks is determined from the statistical data during the query phase. Furthermore, effective complexity management methods are devised by adjusting the decoding path's length or refining constraints on the involved blocks. Thirdly, a simple auxiliary criterion is introduced to reduce computational complexity by minimizing the number of candidate codewords before selecting the optimal estimate. Extensive experimental results and complexity analysis strongly support the proposed framework, demonstrating its advantages in terms of high throughput, low complexity, independence from noise variance, in addition to superior decoding performance.
\end{abstract}



%% file: sections/introduction.tex
\section{Introduction}
\label{intro_sec}
Channel coding plays a critical role in ensuring reliable information transmission within telecommunication systems. Among linear block codes, such as low-density parity-check (LDPC) codes by Gallager \cite{gallager62}, they have distinguished themselves for their remarkable error correction capabilities, even approaching the theoretical Shannon limit asymptotically \cite{mackay96}. In domains like the Internet of things (IoT), where low complexity and minimal latency are paramount, standard iterative belief propagation (BP) decoders of LDPC codes may resort to approximations like min-sum (MS) variants due to computational constraints \cite{zhao05,jiang06}.
In contrast to optimal maximum likelihood (ML) decoding, BP proves suboptimal for finite-length LDPC codes due to the presence of short cycles in the code structures. These simplified MS versions exacerbate this limitation, driving significant efforts towards narrowing this performance gap.

On the other hand, due to their inherently large minimum distance, classical short codes, such as extended BCH codes or RS codes, have reignited academic interest for their potentially superior error correction capability, aligning well with the stringent demands of ultra-reliable and low-latency communication standards like the sixth generation (6G) or IoT \cite{yue2023efficient}. Among others, ordered statistics decoding (OSD) \cite{Fossorier1995}, known for its universal applicability, offers a theoretical approximation of the ML decoder for short linear block codes but grapples with complexity issues in practice. Consequently, substantial efforts have been directed towards reducing its computational burden while preserving the decoding performance of conventional OSD. In \cite{wu2004new}, a two-stage processing framework was proposed to address the ML decoding challenge, initially identifying a minimum sufficient set, followed by OSD equipped with an effective stopping criterion. In \cite{alnawayseh2012ordered}, compliant with a simplistic prior segmenting design of the most reliable basis (MRB) of OSD, the generation of test error patterns (TEPs) is intentionally constrained to alleviate computational complexity. However, this approach suffers from decoding degradation due to the neglect of potential authentic error patterns not covered by the segmenting design. In the segmentation-discarding decoding algorithm of \cite{yue2019segmentation}, authors recommended customizing each received sequence by dynamically partitioning its entire MRB into several segments, with boundaries determined by the magnitude of bits in MRB. Consequently, the generation of TEPs is managed based on the partition result, achieving a remarkable reduction in complexity by leveraging discarding and early stopping rules to avoid futile TEPs checking. In \cite{yue2021probability}, a probability-based OSD variant was devised to mitigate complexity by checking TEPs using two floating thresholds per received sequence. This approach, validated as one of the state-of-the-art (SOTA) decoders, preserves performance while reducing complexity effectively. However, the precision requirement in various probability computations of TEPs may pose challenges for the necessary quantization operations in practice. In \cite{choi2019fast}, a similar threshold technique is employed to alleviate intense complexity. Essentially following conventional order-$p$ OSD procedures, it requires threshold checking before any reprocessing phase of OSD. As a result, early termination of decoding triggered by the checking failure can save the dominant search complexity of subsequent OSD reprocessing phases, reported no perceivable performance loss for BCH codes. In \cite{cavarec2020learning}, two sufficient and necessary conditions were integrated into OSD when traversing its codeword candidates, facilitating the decoding process to prune out the invalid search of TEPs.

Not limited to reliability information of MRB, \cite{valembois2004box,yang2022reduced} take into account the fact that the reliability information of bits in the least reliable basis (LRB) but adjacent to the boundary of LRB and MRB are also sufficiently reliable to impose additional constraints on the incoming TEPs to shrink the search space of codeword candidates. In \cite{liang2022low}, based on the partial constraints specified by the local parity check matrix, the TEPs are generated according to the existing serial list Viterbi algorithm, enabling the early stopping criteria to take effect with minimal performance loss. In \cite{yue2022linear}, extra constraints on the codeword candidates generation are imposed, specifying that the optimal codeword estimate has to be one of the feasible solutions of a system of linear equations derived from the highly reliable bits information in LRB. Interestingly, \cite{choi2020fast} observes that the partial syndrome information derived from MRB should align better in the significant part of LRB than the insignificant part of it, expediting the decoding by saving the re-encoding and quite a few real additions of the unpromising TEPs.

Furthermore, when the computational complexity of Gaussian elimination (GE) invoked in reducing the parity check matrix $\mathbf{H}$ into its systematic form becomes significant in high signal-to-noise ratio (SNR) regimes, \cite{choi2021fast} suggests utilizing several prefabricated templates to generate candidate codewords for the current received sequence, aiming to save operations involved in GE. Additionally, \cite{yue2022ordered} presents two decoding testing conditions, in the form of probability estimation, to support the decision of whether to conduct the OSD or not via threshold checking. However, the effectiveness of both methods was observed primarily in sufficiently high SNR regimes.

Theoretically, in \cite{yue2021revisit}, the Hamming distance and weighted Hamming distance distributions of the OSD were thoroughly analyzed from the perspective of ordered statistics. As a result, many solid insights can be derived to assist in considerably reducing complexity while retaining decoding performance. As an extension of \cite{liang2022low}, after approximating the local parity-check matrix as a totally random matrix, the authors of \cite{liang2024random} derived the performance upper bound for the specified decoding method by exploiting the random coding approach.

While \cite{tataria20216g} provides a global outline of mobile communications development, interested readers can refer to \cite{shirvanimoghaddam2018short,yue2023efficient} for a comprehensive overview of SOTA channel coding techniques, including the latest advancements and the confronted challenges.

For LDPC code decoding, integrating lower-order OSD as a reprocessing step within the iterative BP framework has been proposed \cite{Fossorier1999,Fossorier2001}. However, such integration would largely negate the advantages of parallelizable BP implementation crucial for achieving high decoding throughput due to the inherently serial nature of OSD. To address this capacity mismatch, \cite{jiang2007reliability} suggested employing OSD as a post-processing remedy specifically for BP decoding failures. In this approach, the weighted sum of soft information per BP iteration results in an improved bit reliability measurement beneficial for OSD. Recognizing the lower computational complexity of normalized min-sum (NMS) \cite{zhao05} compared to OSD, \cite{baldi2016use} elaborated on an NMS-OSD hybrid framework for short LDPC codes. In this framework, NMS handles decoding while OSD focuses on further improving decoding by addressing received sequences that failed NMS decoding. This relayed mode effectively alleviates the bottleneck concern between concatenated parts, as the decoding of parallelizable NMS typically leaves few failures for OSD post-processing in the relevant SNR regimes.
Under the assumption that bit reliability deteriorates further after the last iteration of BP decoding failure, \cite{zhang2023efficient} proposes conducting several BP-like iterations on the received sequence to enhance bit reliability before forwarding it to OSD. However, this method relies on prior knowledge of the noise variance, and inaccurate estimation of it may result in performance degradation.

In the past decade, the emerging field of deep learning has shown tremendous promise across various domains, including image processing \cite{razzak18}, natural language processing \cite{young18}, autonomous driving \cite{grigorescu20}, and, more recently, error correction coding. Notably, Nachmani et al.'s pioneering work \cite{nachmani16} proposed directly unrolling standard BP iterations into a dedicated neural network (NN) with trainable parameters for each network edge, effectively addressing the 'curse of dimensionality' in code space \cite{gruber17} by adapting existing decoding methods. Without prior structural considerations, it is challenging for an NN to effectively learn the inherent code structure from scratch. To simplify the complexity associated with tanh or inverse tanh functions in standard BP, NN adaptations of NMS and offset min-sum (OMS) decoders \cite{jiang06} have demonstrated remarkable decoding performance for some classical block codes \cite{nachmani18, liang18, lugosch18, lugosch18-1xVPf, wang20, helmling19}. An intriguing approach involves utilizing a senior NN as a teacher to guide a new NN for enhanced decoding through training \cite{nachmani22}. Furthermore, in a study by Buchberger et al. \cite{buchberger20}, an NN was developed to identify and exclude extraneous check nodes from the overcomplete parity check matrix $\mathbf{H}$ in each iteration, achieving a commendable balance between performance and complexity for short LDPC codes. However, extending this approach to longer codes is challenging due to the nontrivial task of finding a suitable overcomplete $\mathbf{H}$. In another investigation \cite{buchberger21}, the integration of an NN to determine which variable nodes to decimate resulted in a neural BP with decimation (NBP-D) scheme, which effectively ensemble multiple decoding results and substantially boosts decoding performance, despite the marked increase in computational and implementation complexity. Addressing decoding failures in BP caused by the dominant absorbing sets of the LDPC code structure, a suite of recurrent neural network (RNN) models was tailored and individually trained for each dominant absorbing set \cite{Rosseel2022}. The integration of the BP-RNN ensemble and OSD was found to closely approach ML performance for short LDPC codes. However, identifying the absorbing sets and effectively training these RNN models requires considerable effort, in addition to their expensive implementation complexity. Moreover, the decoding throughput is influenced by the intertwined processing mode between the RNN models and OSD, considering the similar capacity mismatch as mentioned earlier.

Regarding the OSD variants alone, most of the schemes mentioned earlier are based on the independently and identically distributed (i.i.d) assumption for additive white Gaussian noise (AWGN) corrupting the received sequences. However, when addressing NMS decoding failures, these schemes need to be adapted, and the resulting validity remains to be confirmed. Furthermore, most schemes explicitly or implicitly depend on acquiring noise variance as a precondition. For example, in \cite{yue2021probability}, all probability computations cannot proceed without knowledge of the noise variance, while \cite{choi2019fast} requires this knowledge in evaluating a specific hyperparameter in the threshold term. In addition, most of them are conditioned on a case-by-case study for the received sequences to achieve practically optimal decoding with minimal complexity. As a result, the serial processing mode resulting from on-the-fly determination of the TEP processing sequence ruins any parallel variants, thus severely impacting the decoding throughput. In contrast, conventional order-$p$ OSD is optimal in terms of decoding performance for its traversal of all TEPs and enjoys the convenience of parallel processing of TEPs for its predetermined debut order, whose drawback lies in the surging computational complexity with the increased $p$.

To address these issues, we corroborate the similar hybrid framework of \cite{baldi2016use}, which was highlighted for achieving a good tradeoff between several key factors such as performance, decoding complexity, and low latency. However, instead of the general option of utilizing the magnitude presented in the initial or the last iteration of NMS decoding as a reliability measure for the OSD, we forge a new reliability measurement for the codeword bits in the hope of maximizing the efficacy of OSD. Differing from the case-by-case study strategy, we suggest a parallelizable OSD implementation where a fixed number of TEPs are checked in a fixed order for any received sequence, effectively expediting the decoding process at the cost of a mild increase in computational complexity.
The main contributions of this paper are as follows:

\begin{itemize}
    \item[*] A novel approach called decoding information aggregation (DIA) is introduced to forge a bit reliability measurement by distilling information from the iterative trajectory of decoding failures. Implemented via a simple CNN model, DIA leverages bitwise soft information collected from all iterations of a posteriori logarithmic likelihood ratios (LLRs) during NMS decoding. This technique effectively enhances OSD decoding and shows promising benefits for various OSD variants in handling iterative decoder failures.
    
    \item[*] To address the inefficiency of conventional OSD caused by exhaustive searching while preserving its parallel processing benefits, we proposed a parallelizable OSD to decode NMS decoding failures. This approach follows a decoding path composed of a list of blocks, named order patterns, whose priorities are determined empirically. Specifically, the MRB is partitioned into several uneven segments, each of which is assigned a labeled Hamming weight to regulate the number of non-zero elements within itself. Then, the order patterns distinguish themselves by the included TEPs that qualify for the segment labeling and are prioritized via counting the number of trappings of authentic error patterns in the query phase after sufficient sampling. In the test phase, the decoding path can be readily adapted to meet various scenario requirements by tweaking its length or imposing more constraints on the order patterns.
    
    \item[*] An auxiliary criterion is proposed involving simple integer operations to threshold promising codeword candidates. It can significantly reduce the list size of codeword candidates with minimal performance loss after fine-tuned evaluation of thresholds; nearly half of the real sums requested when computing the optimal codeword estimate were saved in simulations.
    
    \item[*] For short LDPC codes, the unified framework of NMS-DIA-OSD demonstrated comparable performance to the SOTA decoders experimentally, while featuring the characteristics of high throughput, low complexity, and no need for noise variance, etc. However, it was also revealed that the challenges confronting the existing schemes remain maximal for the unified framework with respect to longer LDPC codes, and DIA application for classical BCH codes yielded no satisfactory results due to the absence of effective iterative message passing.
\end{itemize}

The rest of the paper is organized as follows: Section \ref{preliminary} provides essential preliminaries about BP decoding variants and OSD variants. In Section \ref{motivation}, the motivations behind our work are elaborated. Section \ref{simulations} demonstrates the experimental results and complexity analysis. Finally, Section \ref{conclusions} concludes the paper with remarks and suggestions for further research.

%% file: sections/preliminary.tex
\section{Preliminaries}
\label{preliminary}

Given a binary row message vector $\mathbf{m} = [m_i]_1^K$, the encoder encodes it into $\mathbf{c} = [c_i]_1^N$ by $\mathbf{c} = \mathbf{m}\mathbf{G}$ in GF(2), where $K$ and $N$ are the lengths of the message and codeword respectively, and the generator matrix $\mathbf{G}$ is assumed to be full-rank without loss of generality.

After the simple binary phase shift keying (BPSK) modulation maps each codeword bit $c_i$ into one symbol $s_i = 1 - 2c_i$, then a sequence $\mathbf{y} = [y_i]_1^N$ with $y_i = s_i + n_i$ is received at the channel output, where $n_i$ denotes the channel noise with mean zero and variance $\sigma^2$. 

Suppose the equiprobability of sending symbols, it is derived the LLR of the $i$-th bit is
\begin{equation}
l_i = \log \left( \frac{{p(y_i|c_i = 0)}}{{p(y_i|c_i = 1)}} \right) = \frac{{2y_i}}{{\sigma^2}}. 
\label{llr_definition}
\end{equation}
Evidently, the magnitude of $y_i$ serves as a reliable metric naturally for assessing the confidence in the $i$-th bit's hard decision $\hat{c}_i$, which is either '1' or '0' depending on the sign of $y_i$. Specifically, $\hat{c}_i = \mathbf{1}(y_i < 0)$ with $\mathbf{1}(\cdot)$ being the indicator function.

\subsection{BP, MS Variants and Neural Versions}
The bipartite Tanner graph of a code, determined by its parity check matrix $\mathbf{H}$, consists of $N$ variable nodes and $M$ check nodes connected by the edges for the non-zero elements in the $i$-th row and $j$-th column of $\mathbf{H}$ where $i\in \{1,2,\cdots,M\}$ and $j\in \{1,2,\cdots,N\}$. For simplicity, it is assumed $\mathbf{H}$ is full row rank and thus $M=N-K$.

For LDPC codes, standard BP is a competitive decoding scheme, known as optimal under the assumption of a tree-like Tanner graph without cycles. However, for the finite-length codes, due to the inherent short cycles in its $\mathbf{H}$, the decoder degrades into suboptimal as the penalty of  breaking the assumption.

To be self-contained, let us take a close view of the BP flooding messages schedule which facilitates the parallel processing mode. Suppose $t \in \{1,2,\ldots,T\}$ with $T$ being the maximum number of iterations, then the message from variable node $v_i$ to check node $c_j$ at the $t$-th iteration and the message right in the reversed direction are given by \eqref{eq_v2c}\eqref{eq_c2v} respectively:

\begin{equation}
x_{v_i \to c_j}^{(t)} = l_i + \sum_{\substack{c_p \to v_i\\p \in \mathcal{C}(i)/j}} x_{c_p \to v_i}^{(t - 1)}
\label{eq_v2c}
\end{equation}

\begin{equation}
x_{c_j \to v_i}^{(t)} = 2\tanh^{-1}\left( \prod_{\substack{v_q \to c_j\\ q \in \mathcal{V}(j)/i}} \tanh\left( \frac{x_{v_q \to c_j}^{(t)}}{2} \right) \right)
\label{eq_c2v}
\end{equation}

where $\mathcal{C}(i)/j$ denotes the set of all neighboring check nodes of $v_i$ excluding $c_j$, and $\mathcal{V}(j)/i$ the set of  all neighboring variable nodes of $c_j$ excluding $v_i$. And all $x_{c_p \to v_i}^{(0)}$ are initialized to be zeros.

The alternating of \eqref{eq_v2c}\eqref{eq_c2v} is regarded as one BP iteration, and the process continues before reaching the $T$-th iteration. Meanwhile, the a posteriori LLR of all bits, defined as \eqref{eq_bit_decision}, is computed so that the hard decision at the $t$-th iteration which align with the resultant signs, is attempted to check whether or not the all-zero syndrome is satisfied to trigger early termination of the decoding. 

\begin{equation}
x_{v_i}^{(t)} = l_i + \sum_{\substack{c_p \to v_i\\ p \in \mathcal{C}(i)}} x_{c_p \to v_i}^{(t - 1)}
\label{eq_bit_decision}
\end{equation}

To reduce intensive computation, the MS scheme
was proposed to substitute a simple approximation \eqref{eq_ms} for the computation of $\tanh$ or $\tanh^{-1}$ functions in \eqref{eq_c2v}, at the cost of some performance loss:

\begin{equation}
x_{c_j \to v_i}^{(t)} = s_{c_j \to v_i}^{(t)} \omega_{c_j \to v_i}^{(t)}
\label{eq_ms}
\end{equation}
\begin{align*}
s_{c_j \to v_i}^{(t)} &= \prod_{\substack{v_q \to c_j\\ q \in \mathcal{V}(j)/i}} \text{sgn}\left( x_{v_q \to c_j}^{(t)} \right) \\
\omega_{c_j \to v_i}^{(t)} &= \min_{\substack{v_q \to c_j\\ q \in \mathcal{V}(j)/i}} \left| x_{v_q \to c_j}^{(t)} \right|
\end{align*}
where sgn($\cdot$) denotes the sign function.

As opposed to standard BP whose decoding performance may suffer from inaccurate estimation of channel noise $\sigma^2$ \cite{lugosch18-1xVPf}, the MS demonstrates another attractiveness  scale invariance to $\sigma^2$. That is, no impact on performance is expected for the MS in case of any scaling of received sequences. 

To narrow the performance gap between MS and BP decoders, NMS or OMS were proposed to appropriately weight or offset the min term $\omega_{c_j \to v_i}^{(t)}$ in \eqref{eq_ms}.

BP, NMS, or OMS can be readily transformed into a trellis-like NN by unrolling each iteration of the decoding process, with the added complexity of imposing weights or offsets on the neural network edges. Among the others, NBP has the highest complexity, and neural OMS generally lags behind neural NMS  in terms of decoding performance \cite{lugosch18-1xVPf}. Therefore, we opt to the neural NMS and favor the simplest form of it, which is seemingly the original NMS itself. And the annotation is all the NN edges related to the min term $\omega_{c_j \to v_i}^{(t)}$ in \eqref{eq_ms} are all weighted by a shared parameter $\alpha$ across iterations.

in  training phase, the neural decoder is supplied with a batch of data and assessed for its output using the standard cross-entropy loss function. Following this, a stochastic gradient descent (SGD) optimizer is employed to adjust the single parameter evaluation. The training process iterates until a stopping criterion is met, such as when the fluctuation of the loss value over several consecutive iterations falls within a predefined range. It has been observed that a well-trained NMS can effectively bridge most of the performance gap between the MS and the BP. 

\subsection{OSD of Linear Block Codes}
\label{convention_osd}

The OSD variants of linear block codes fall into two main categories: $\mathbf{G}$-oriented and $\mathbf{H}$-oriented. We specifically focus on $\mathbf{H}$-oriented OSDs in this paper due to their procedural duality with $\mathbf{G}$-oriented OSDs, as well as their de facto performance and complexity. Another justification is that the $\mathbf{H}$ matrix of LDPC codes is sparse, unlike their $\mathbf{G}$ matrix, making the former less demanding in terms of computation for GE operations.

For any $\mathbf{H}$-oriented OSD variants, it is crucial to note that each bit position of a received sequence $\mathbf{y}=\mathbf{y}^{(0)}$ is rigidly linked with each column position of the initial matrix $ \mathbf{H}=\mathbf{H}^{(0)} $. This implies two latent interactions between the sequence $\mathbf{y}^{(i)}$ and the matrix $\mathbf{H}^{(i)}$ following the later OSD procedures:

\begin{enumerate}[label=(\roman*)]
    \item Permutation of all (or some) bits of $\mathbf{y}^{(i-1)}$ into $\mathbf{y}^{(i)}$ leads to the rearrangement of all (or some) columns of $\mathbf{H}^{(i-1)}$ into $\mathbf{H}^{(i)}$. \label{itemone}
    \item Requested column swapping in GE of $\mathbf{H}^{(i-1)}$ into $\mathbf{H}^{(i)}$ results in the rearrangement of the involved bits of $\mathbf{y}^{(i-1)}$ into $\mathbf{y}^{(i)}$. \label{itemtwo}
\end{enumerate}

For conventional order-$p$ OSD, the allowed Hamming weight of all its TEPs is at most $p$. Essentially, it involves searching for the optimal estimate within reasonable complexity. The procedures of conventional order-$p$ OSD are unfolded as follows:

\begin{enumerate}
    \item Sort $\mathbf{y}^{(0)} $ in ascending order of reliability measurement (commonly the magnitude) to obtain $\mathbf{y}^{(1)}$. Meanwhile, $\mathbf{H}^{(0)}\mapsto \mathbf{H}^{(1)}$ due to the permutation $\mathbf{y}^{(0)}\mapsto \mathbf{y}^{(1)}$ in interaction \ref{itemone}. 
    
    \item The GE operation, including several elementary row operations and latent column swaps, reduces $\mathbf{H}^{(1)}$ to its systematic form $\mathbf{H}^{(2)} = [\mathbf{I}:\mathbf{Q}_2]$. Simultaneously, $\mathbf{y}^{(1)}\mapsto \mathbf{y}^{(2)}$ is triggered by $\mathbf{H}^{(1)}\mapsto \mathbf{H}^{(2)}$ in interaction \ref{itemtwo}. Then the resultant $\mathbf{y}^{(2)}$ is segmented into two parts: the MRB consisting of the tail $K$ bit positions of the sorted bits, and the LRB for the remaining bit positions. Notably, some references favor the term "most reliable independent positions" (MRIPs) to address both the independence and reliability properties of interested bits, while the adopted MRB notation in this paper actually means exactly the same as the MRIPs. Additionally, we emphasize that the elementary row operation on $\mathbf{H}^{(1)}$ has no impact on the permutation of $\mathbf{y}^{(1)}$ for its effect equal to left multiplication of $\mathbf{H}^{(1)}$ by some matrix.
    
    \item The hard decision of $\mathbf{y}^{(2)}$ on bits belonging to the MRB yields the anchoring point $\mathbf{c}_a$, on which the qualified TEPs will be superimposed in XOR mode so as to list all the candidates. Specifically, for each TEP $\mathbf{e}_j = [e_i]_1^K$, $j \in \{1,2,\ldots,\sum_{i=0}^p \binom{K}{i}\}$, $\overline{\mathbf{c}}_{j,2} = \mathbf{c}_a \oplus \mathbf{e}_j$ serves as the $j$-th estimate of the MRB. Due to the parity check constraints of $\mathbf{H}^{(2)}{\overline{\mathbf{c}}_j^\top} = \mathbf{0}$, a codeword candidate $\overline{\mathbf{c}}_j = [\overline{\mathbf{c}}_{j,1} : \overline{\mathbf{c}}_{j,2}]$ is secured, where $\overline{\mathbf{c}}_{j,1} = \overline{\mathbf{c}}_{j,2}\mathbf{Q}_2^\top$ claims the $j$-th estimate of the LRB part.
    
    \item All codeword candidates $\overline{\mathbf{c}}_j$, $j \in \{1,2,\ldots,\sum_{i=0}^p \binom{K}{i}\}\}$, compete for the optimal estimate under the criterion:

\begin{equation}
\label{argmin_dis}
\overline{\mathbf{c}} = \mathop {\arg\min }\limits_{\overline{\mathbf{c}}_j = [c_i]_1^N} \sum\limits_{i = 1}^N {\mathbf{1}(\check{c}_i \ne {c}_i)\left| {{y}^{(2)}_i} \right|}
\end{equation}

where $\check{c}_i$ is the hard decision of $y^{(2)}_i$ (the $i$-th entry of $\mathbf{y}^{(2)}$).

    \item The estimated codeword $\hat{\mathbf{c}}$ is identified after reversing all the involved bit swaps for $\overline{\mathbf{c}}$.
    $\hspace{1cm}<\mathbf{end}>$
\end{enumerate}

%% file: sections/motivation.tex
\section{Motivations}
\label{motivation}
\subsection{DIA Model}
\subsubsection{Rationale of DIA}
In the realm of OSD, the magnitude of soft information from received codeword bits is conventionally treated as the reliability metric. However, our exploration into LDPC codes revealed a potential avenue for enhancing this metric, even in scenarios where NMS decoding fails to converge.

Specifically, we scrutinized the reliability assessment of each bit, whether based on the initial $|{y}_i^{(0)}|$ or the final $|x_{v_i}^{(T)}|$ of the NMS decoding iteration, as detailed in \eqref{eq_bit_decision}. Yet, the availability of a posteriori LLRs for every codeword bit per decoding iteration prompted us to conjecture that harnessing the entire decoding trajectory of these a posteriori LLRs through neural networks could yield a more reliable metric, closely aligning with the ground truth when quantified by cross entropy.

Moreover, within our proposed framework, the OSD operates as a postprocessor. Thus, it becomes reasonable to utilize samples of NMS decoding failures for training neural networks. This intuition propels us to validate our hypothesis using CNN models, renowned for their quick responsiveness and manageable complexity. It's worth noting that in this context, CNN and DIA are essentially interchangeable, although the latter emphasizes its role in decoding information aggregation.

Recognizing the computational constraints, we acknowledge that the effectiveness of OSD heavily hinges on the meticulous selection of $K$ bit positions to populate the MRB based on sorted reliability measurements. Undoubtedly, any improvement in reliability would significantly enhance the OSD's performance. However, it's crucial to highlight that the reliability metric forged by the DIA model is primarily used for preliminary operations preceding \eqref{argmin_dis}. Our investigations have revealed that in some cases, authentic error patterns remain undiscernible due to failures in \eqref{argmin_dis}. Hence, it's imperative to revert to the magnitude of the original received sequences as the reliability metric when invoking \eqref{argmin_dis}.

An illustrative example is provided on how to implement DIA for the CSDSS LDPC (128,64) code \cite{helmling19}.

\subsubsection{Architecture of DIA Model}
Let $T=12$ for the NMS decoding of the chosen code. With respect to each of its decoding failures, the decoding trajectory of the $i$-th bit is treated as a time series whose length is $T+1=13$, after taking into account the received $y_i$ as well. Thus, the issue of bit reliability estimation is transformed into a time series estimation, and a refined metric is expected to be forged via capturing the inherent modes and tendencies among the data observations.

\begin{figure}[htbp]
\centering
\centerline{\includegraphics[width=0.45\textwidth]{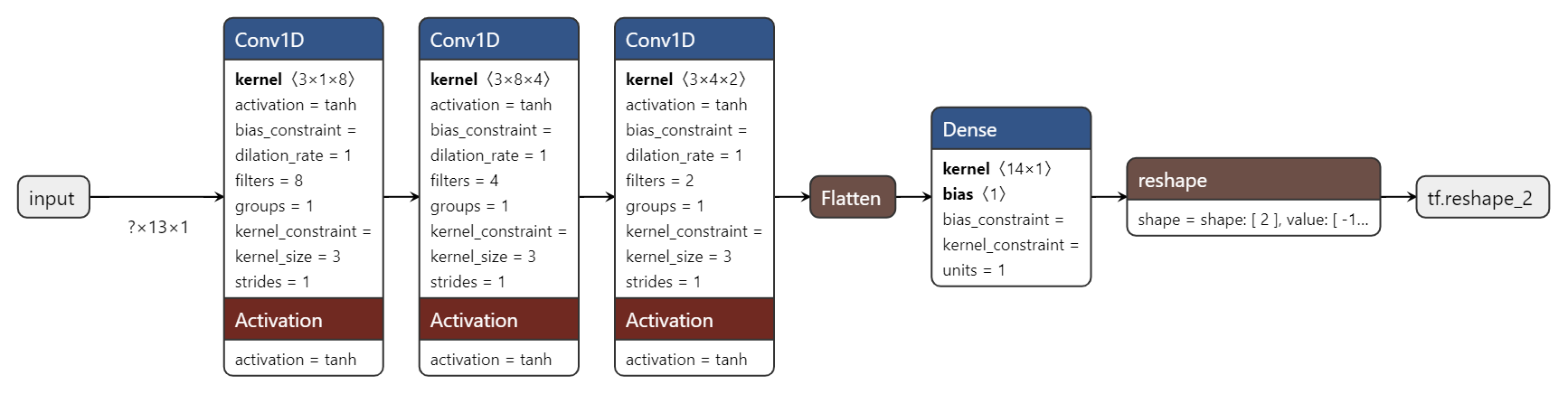}}
\caption{CNN implementation of the DIA model for CCSDS LDPC (128,64) code}
\label{model_cnn_128}
\end{figure}

To address this, a simple CNN architecture is employed, which consists of three 'Conv1D' layers, one 'Flatten' layer, and one 'Dense' layer from the Keras package \cite{chollet2015keras}. These layers are meticulously configured with parameters such as 'kernel$\_$size=3', 'strides=1', 'padding=valid', 'activation=linear' (or 'tanh'), etc., as illustrated in Fig. \ref{model_cnn_128}.
In the architecture, a stack of '3x1' filters convolve over the one-dimensional decoding trajectory individually, extending the receptive field from the current iteration to neighboring iterations to facilitate information fusion. The deeper the concatenated 'Conv1D' layers, the wider the receptive field grows. With the aid of the 'Flatten' layer, which supports adaptation to varied $T$ settings, the 'Dense' output layer merges the available information to yield a new reliability estimate about the current bit. For simplicity, the number of filters is configured to be 8, 4, and 2 for the three 'Conv1D' layers in the model separately for any code, but retraining is needed for each code under consideration.

\subsubsection{Training Logistics of DIA Model}
The training data is obtained from the recorded a posteriori LLRs per iteration for all decoding failures in the NMS decoding at SNR=2.7dB for this code and serves as the sole source to feed the DIA model, unless stated otherwise. In the mini-batch training mode with the input data matrix of shape $N \times (T+1)$, each row represents a time series, and the batch size is equal to the number of rows of the matrix, which is identical to the number of codeword bits in our setting. The Adam optimizer \cite{kingma14} is utilized  to update the trainable parameters and minimize the common cross-entropy loss function \eqref{ce_loss}, defined as follows:
\begin{equation}
\label{ce_loss}
\ell_{ce} (\mathbf{c},\widehat{\mathbf{c}}) = \sum_{i=1}^{N} \sum_{j=0}^1 \left( p(c_i = j) \cdot \log \frac{1}{p(\hat{c}_i = j)} \right)
\end{equation}
where $p(c_i = j)$ takes the value of 0 or 1 depending on the ground truth, and $p(\hat{c}_i = j)$ denotes the derived probability from the LLR output of the DIA for the $i$-th bit. In particular, considering that $\sigma^2$ is known in the training phase, the probability and its corresponding a posteriori LLR are mutually determined. While the cross-entropy evaluation is no longer necessary in the testing phase, the DIA output exclusively serves as the reliability metric for sorting the codeword bits in the post-processing of OSD.

Given the simple model complexity, we can safely set the initial learning rate to $0.01$, and the training process will rapidly reach its stable state within $1000$ steps, requiring only a few minutes on a personal computer with the '2.60GHz Intel i7-6700HQ' processor.

\subsubsection{Impact of DIA on NMS Decoding Failures}

\begin{figure}[htbp]
    \centering
    \begin{subfigure}{0.24\textwidth}
     \captionsetup{font=small} 
        \includegraphics[width=\linewidth]{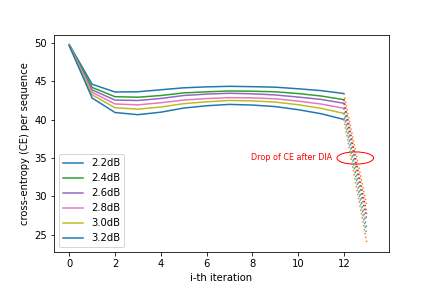}
        \caption{Cross-Entropy Evolution}
        \label{cross_entropy_128}
    \end{subfigure}
    \hfill
    \begin{subfigure}{0.24\textwidth}
        \captionsetup{font=small} 
        \includegraphics[width=\linewidth]{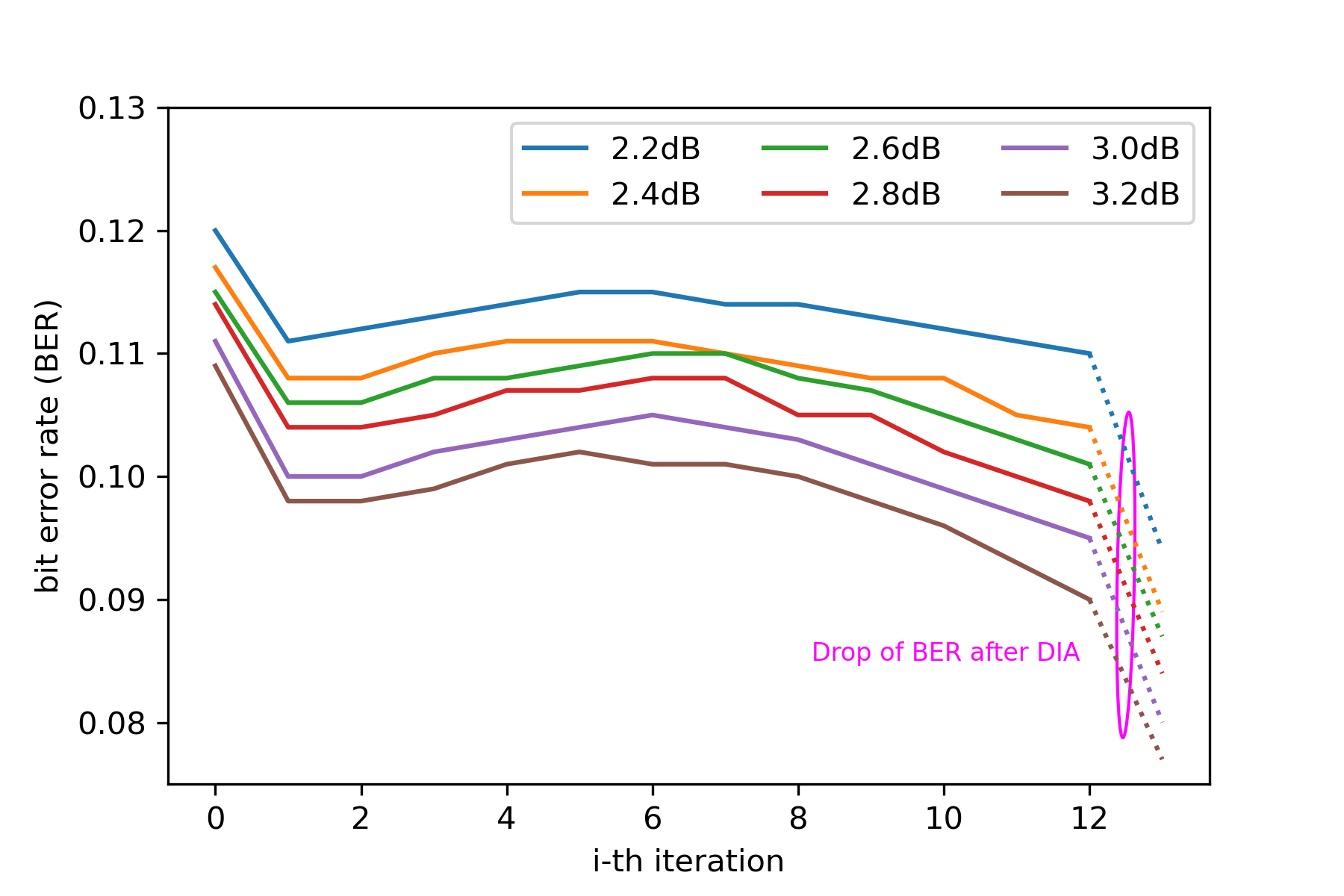}
        \caption{BER Evolution}
        \label{ber_iteration128}
    \end{subfigure}
    \caption{Comparison of cross-entropy and BER evaluation per decoding iteration for NMS ($\alpha=0.78$) decoding failures of LDPC (128,64) code with $T=12$. The plot includes the outputs of the DIA model, trained at SNR = 2.7dB and applied to all SNR points.}
\label{cross_entropy__ber_plot128}
\end{figure}

Regarding NMS decoding failures, we can monitor the fluctuation of cross-entropy throughout each decoding iteration and compare it with the output of the DIA model for validation.

As shown in Fig.~\ref{cross_entropy__ber_plot128}, several observations emerge: Firstly, the decrease in cross-entropy reaches a saturation state rapidly after the initial iterations, coinciding with a similar trend observed in the bit error rate (BER). This suggests limited potential for improving bit reliability even with extended decoding iterations. Conversely, the sharp decline in cross-entropy evaluation for DIA output strongly implies its ability to effectively enhance the original reliability measurement of correct bits while suppressing that of erroneous bits on average. This aligns with the demand of OSD to prioritize correct bits with the largest magnitudes to fill its MRB. Secondly, the cross-entropy curves for different SNRs are closely aligned, as are the BER curves, suggesting that a model trained at a specific SNR may be effective across all SNR values. Lastly, the BER curves generally fluctuate and end with a mild drop due to the hard-decision output of DIA, showcasing its corrective ability.

\begin{figure}[htbp]
    \centering
    \begin{subfigure}{0.24\textwidth}
        \captionsetup{font=small} 
        \includegraphics[width=\linewidth]{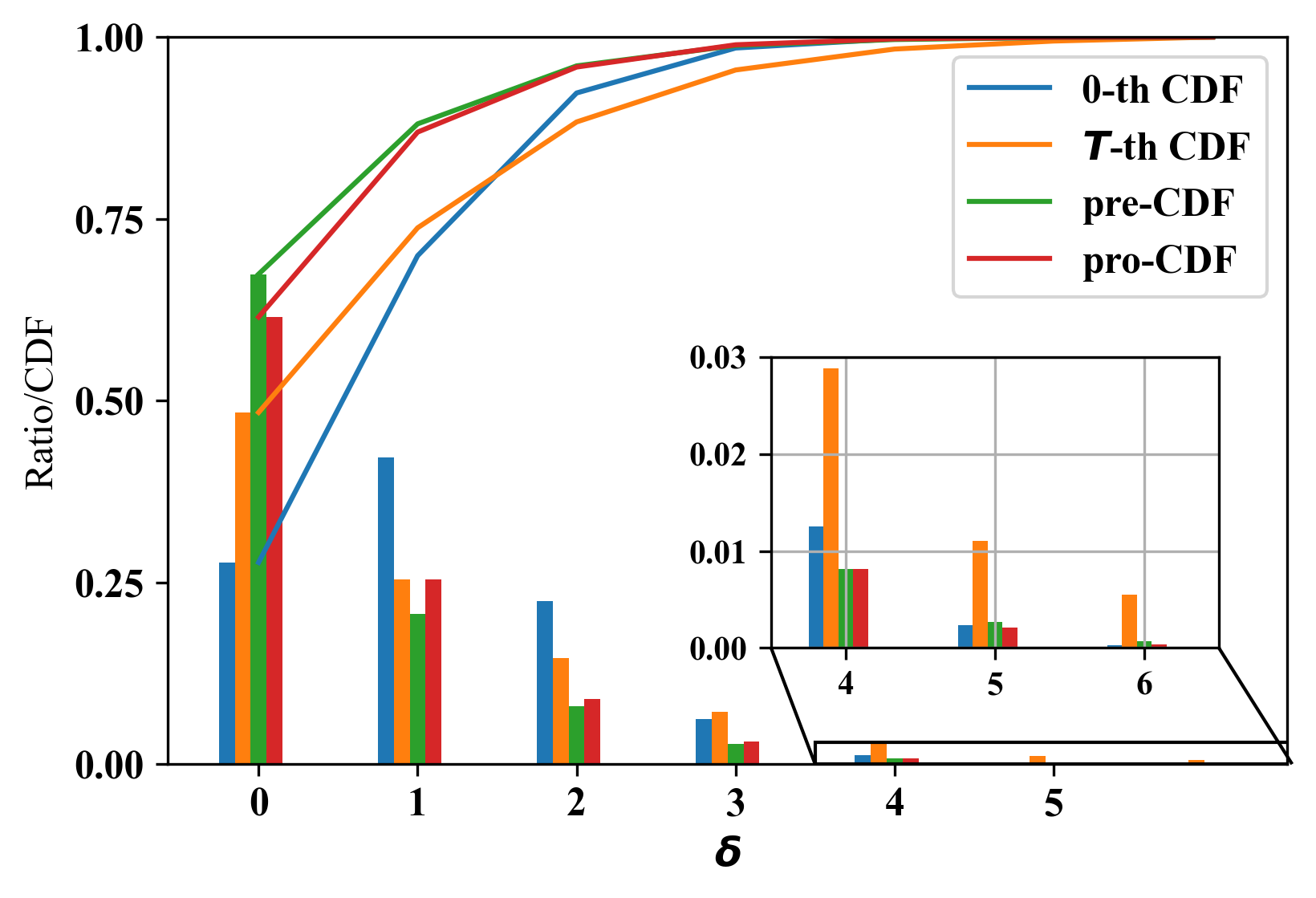}
        \caption{CDF of $\delta$ at SNR=2.2dB}
        \label{delta_2dot2_128}
    \end{subfigure}
    \hfill
    \begin{subfigure}{0.24\textwidth}
        \captionsetup{font=small} 
        \includegraphics[width=\linewidth]{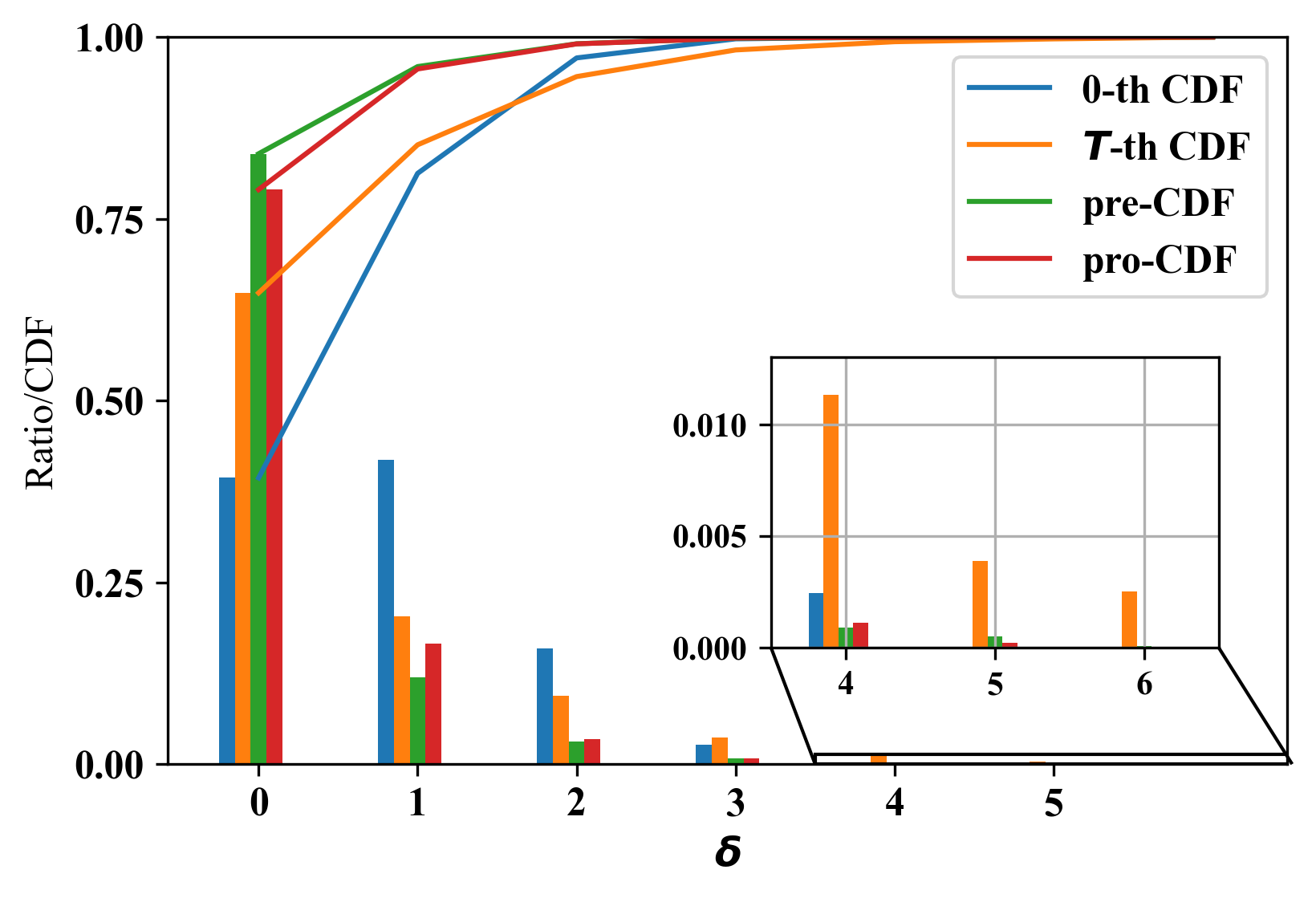}
        \caption{CDF of $\delta$ at SNR=3.2dB}
        \label{delta_3dot2_128}
    \end{subfigure}
    \caption{Ratios and cumulative distribution functions (CDFs) of the number of erroneous bits $\delta$ in the MRB of NMS decoding failures for LDPC (128,64) code (the cases of $\delta\ge6$ merged into $\delta=6$ for convenience of plotting)}
    \label{order_distribution_128_2dot2_3dot2}
\end{figure}

Despite the slight decrease in BER when comparing $T=0$ with $T=12$ as shown in \ref{ber_iteration128}, a posteriori LLRs of the $T$-th iteration were preferred over the initial LLRs of the received sequences as the reliability metric \cite{baldi2016use,zhang2023efficient}. This choice is made in case the occurrence of either refusing correct bits of low magnitude or admitting erroneous bits of high magnitude into the MRB happens, justified partially by the zoomed ratios in Fig.~\ref{order_distribution_128_2dot2_3dot2} where the $T$-th iteration's performance is the worst in terms of the number of erroneous bits $\delta\ge4$ in the MRB. In comparison, the DIA model addresses this issue by not only reducing the number of erroneous bits in the MRB but also elevating the ranks of correct bits in sorting within the MRB. The latter ensures that if erroneous bits do enter the MRB, they are more likely to be confined to a narrower area of the sorted MRB, rather than scattered throughout the entire MRB. This concentration effect can be leveraged in the design of our OSD variant, as validated by extensive simulations in Section \ref{simulations}.

Regarding the former claim, validation can be achieved by examining the cumulative distribution function (CDF) regarding $\delta$. As depicted in Fig.~\ref{order_distribution_128_2dot2_3dot2}, whether at low SNR=2.2dB or high SNR=3.2dB, the DIA model is reinforced by the observation that the 'pre-CDF' curve consistently hovers over that of '0-th CDF' or '$T$-th CDF'. Here, 'pre-CDF' denotes the MRB hard-decision after sorting with the aid of DIA but before GE operation, while '0-th CDF' and '$T$-th CDF' represent the cases of initial LLRs and a posteriori LLRs of the last iteration of NMS decoding, respectively. Additionally, it's notable that the curve 'pro-CDF', representing the CDF after GE operation, exhibits inferior performance compared to 'pre-CDF' due to the side-effects of GE operation. Furthermore, the timing of the confluence of two curves occurs at $\delta_t=1$, extended to $\delta_t\ge2$ for longer codes than the (128,64) code after preliminary investigation. This implies that an OSD with its order $p<\delta_t$ will be negatively impacted by the GE operation.

\subsection{Decoding path-guided adaptive OSD}
While the conventional order-$p$ OSD theoretically promises ML decoding when $p\geq \left \lceil d_{\text{min}}/4-1 \right \rceil$, where $d_{\text{min}}$ represents the minimum distance of the code \cite{Fossorier1995}, its practicality is hindered by the rapidly increasing complexity $O(N^p)$ associated with enumerating all codeword candidates and selecting the optimal estimate among them. Efforts to mitigate this complexity while maintaining decoding performance have been extensively explored, as discussed in Section \ref{intro_sec}. The conventional OSD operates exhaustively, which can be inefficient, but its advantages lie in the prearranged ordering of all relevant TEPs for any received sequence. Specifically, by aligning with the Hamming weight of TEPs, it facilitates straightforward hardware implementation and enables parallel processing of these TEPs, thereby significantly enhancing decoding throughput. In contrast, most SOTA OSD variants prioritize complexity reduction while striving to maximize performance, often adopting serial processing modes that dynamically adjust processing schemes based on the magnitudes of received sequences. Therefore, our aim is to design a novel OSD scheme that strikes a better balance between performance, complexity, and processing throughput, with the assistance of DIA. The feasibility of its implementation is empirically confirmed in Section \ref{simulations}.

The effectiveness of SOTA OSD variants partly stems from their ability to skip unpromising TEPs through meticulous evaluation involving intensive computation. In contrast, the conventional OSD scans TEPs consecutively without preference, leading to inefficiency. Integrating the advantages of both strategies into our new scheme involves partitioning all TEPs into blocks, each containing consecutive TEPs. These blocks are then prioritized based on statistical data pertaining to authentic error patterns identified during the query phase. In our framework, each block is termed an order pattern, and the list of order patterns, after prioritization, forms the decoding path. During the test phase, this decoding path serves as a roadmap for guiding the decoding of any received sequences. Importantly, unlike existing strategies, our proposed scheme leverages knowledge distilled from history records, significantly reducing the computational burden associated with SOTA OSD variants. Furthermore, it preserves the parallelizability property of conventional OSD, making it well-suited for high-throughput scenarios. By flexibly adjusting the decoding path component, we can readily achieve a competitive trade-off between performance, complexity, and decoding throughput.

Upon receiving a sequence, the OSD's MRB is established following magnitude sorting and GE operation. Subsequently, the MRB's ordered bits are segregated into $Q$ segments, with $Q=3$ illustrating the process. Triple thresholds $\lambda_{im}, i\in{1,2,3}$ are then assigned to set maximum Hamming weights individually. A list of order patterns is subsequently generated, specified by $\lambda_i, i\in {1,2,3}$, which governs the Hamming weights of the MRB's segments under the threshold constraints $\lambda_i\le\lambda_{im}, i\in {1,2,3}$. Following this, all sequentially distributed TEPs meeting the order pattern labels populate them in order.

The priority of each pattern significantly impacts OSD decoding performance, especially considering that the decoding path is often truncated to meet strict complexity requirements. By considering order patterns that contain the most occurrences of authentic error patterns in the query phase, we observe that they tend to exhibit similar behavior in the testing phase. Therefore, we prioritize these dominant order patterns at the forefront of the decoding path. Importantly, the decoding path is fully determined prior to the commencement of the testing phase. Various methods, including adjusting the length $l_{pt}$ of the decoding path or skipping specific order patterns by imposing additional constraints, can be employed to customize the existing decoding path to suit the dynamic needs of applications.

\begin{figure}[ht!]
    \centering
    \includegraphics[width=0.45\textwidth]{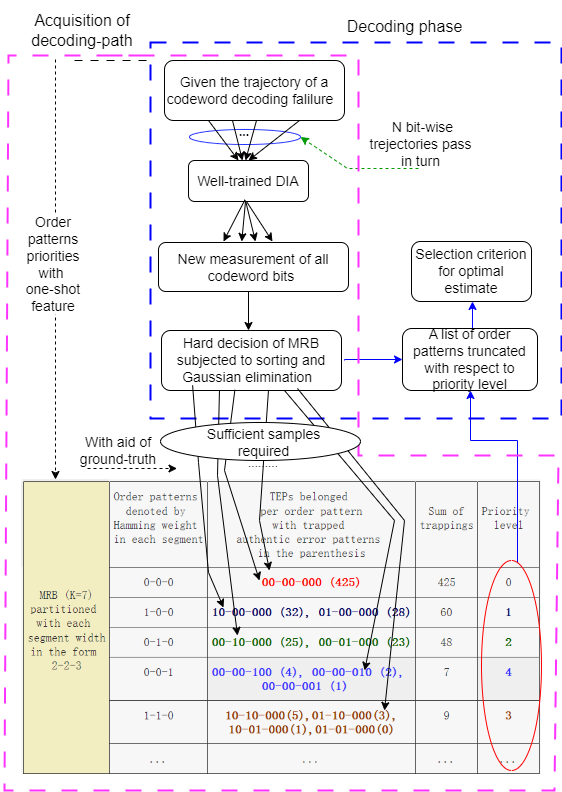}
    \caption{Flow chart depicting the acquisition of the decoding path and its role in OSD decoding, illustrated through a toy example}
    \label{osd_data_flow}
\end{figure}

To better understand the flow chart of the proposed OSD, a toy example is presented in Fig.~\ref{osd_data_flow}, where the acquisition of the decoding path is implemented through the blocks sequentially enclosed in the pink polygon, while the blocks in the blue rectangular present the necessary procedures for OSD decoding.

In general, our alternative opts for a selective searching approach that emphasizes configuring the computational resources on the groups containing the most likely TEPs historically. Two pieces of supportive empirical evidence are presented regarding the statistics of requested swapping counts in GE and order patterns. The former is exploited to offer one clue on how to partition the MRB, and the latter is to validate the feasibility of seeking dominant order patterns.

\subsubsection{Distribution of Requested Swap Counts}

Column swapping commonly occurs when reducing $\mathbf{H}$ to its systematic form. It was theoretically analyzed in \cite{Fossorier1995} assuming i.i.d conditions of received sequences. However, in the case of NMS decoding failures, it is difficult to approximate it probabilistically due to the absence of i.i.d conditions. Therefore, we turn to its empirical evidence. Let $n_{sw}$ denote the occurrences of swaps between the MRB and the LRB, which can help us design a reasonable partition of the MRB that influences the constituents of the decoding path and the ordering of order patterns included.
\begin{figure}[htbp]
    \centering
    \includegraphics[width=0.4\textwidth]{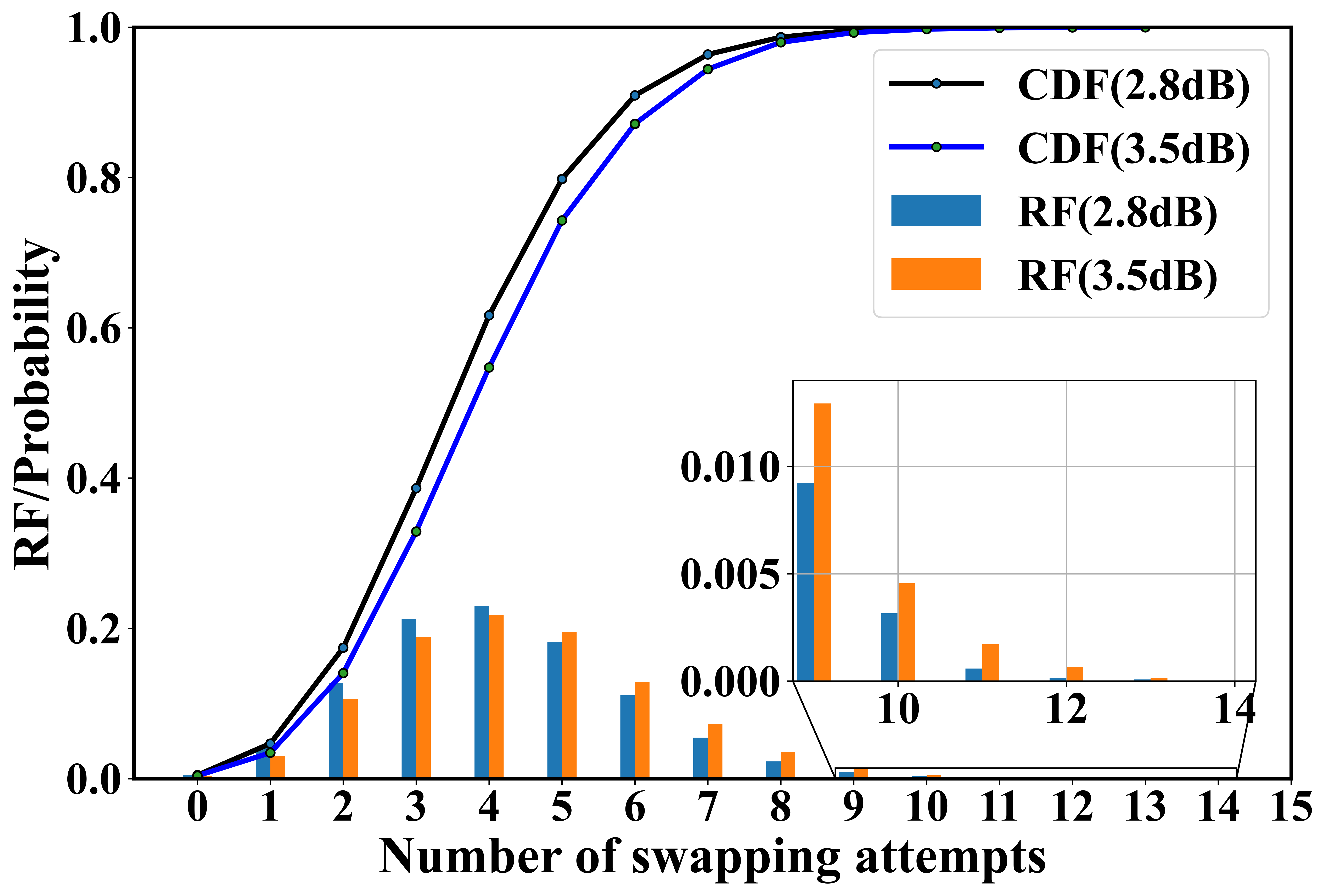}
    \caption{The relative frequency (RF) and CDF of counts of swapping between the MRB and the LRB when reducing $\mathbf{H}$ to its systematic form for decoding failures of $(128,64)$ code at SNR = $2.8, 3.5$dB}
    \label{cdf_128_2dot8_3dot5_swaps}
\end{figure}
As illustrated in Fig.~\ref{cdf_128_2dot8_3dot5_swaps}, the average $n_{sw}$ remains approximately constant at 4 for the $(128,64)$ code, regardless of SNR = 2.8dB or 3.5dB. However, it is expected that the average $n_{sw}$ varies with code block lengths, code rates, and even $\mathbf{H}$ structure. For the (128,64) code, its MRB range is partitioned into $\left \{ 1,2,\cdots,10{\color{red} |}11,12,\cdots 30{\color{red} |}31,32,\cdots  ,64 \right \}$ throughout the simulation. Generally, the delimiting points are closely related to the average $n_{sw}$. For brevity, $\frac{1}{6}N$ and $\frac{1}{2}N$ are used as substitutes for them with respect to all the codes except the (1056,880) code.

\subsubsection{Distribution of Order Patterns}

Once the MRB partition boundaries are determined, the order patterns can be distinguished from each other by the respective labels $[\lambda_1, \lambda_2, \lambda_3]$. During the query phase, each authentic error pattern, also known as ground truth, underlies a decoding failure. Consequently, the counter associated with the order pattern increments by one once the authentic error pattern matches one of the TEPs belonging to the order pattern, and the processing continues until all decoding failures are traversed. Considering the value of all the counters provides us the only reason to prioritize the order patterns, a sufficient sample is demanded to derive a stable evaluation for the available order patterns.

\begin{table}[htbp]
\setlength{\tabcolsep}{1.5mm}
\caption{Statistics about the impact of DIA on distribution of order patterns when identifying the decoding path for the decoding failures of (128,64) code at SNR=3.0dB}
\begin{center}
\resizebox{\linewidth}{!}{
\begin{tabular}{|c|c|c|}
\hline
SNR                     & DIA                                 & Order patterns:ratio                                                                                                                                                                                                                                                                                                                                                                                                                                                                                                                  \\ \hline
                        & No                                  & \begin{tabular}[c]{@{}c@{}}\{'{[}0 0 0{]}': 0.6847, '{[}1 0 0{]}': 0.1452, '{[}2 0 0{]}': 0.0449, '{[}0 1 0{]}': 0.0351, '{[}1 1 0{]}': 0.0269, \\ '{[}2 1 0{]}': 0.0125, '{[}3 0 0{]}': 0.0117, '{[}0 0 1{]}': 0.006, '{[}1 0 1{]}': 0.005, '{[}1 2 0{]}': 0.0049, \\ '{[}0 2 0{]}': 0.0039, '{[}3 1 0{]}': 0.0037, '{[}2 0 1{]}': 0.0028, '{[}4 0 0{]}': 0.0025, '{[}2 2 0{]}': 0.0024,\\  '{[}0 1 1{]}': 0.0024, '{[}1 1 1{]}': 0.0024, '{[}2 1 1{]}': 0.0012, '{[}3 2 0{]}': 0.001, '{[}3 0 1{]}': 0.0009\}\\  Others: 0.0086\end{tabular}                                  \\ \cline{2-3} 
\multirow{-6}{*}{3.0dB} & {\color[HTML]{3531FF} \textbf{Yes}} & {\color[HTML]{3531FF} \textbf{\begin{tabular}[c]{@{}c@{}}\{'{[}0 0 0{]}': 0.8205, '{[}1 0 0{]}': 0.1157, '{[}0 1 0{]}': 0.0215, '{[}2 0 0{]}': 0.0197, '{[}1 1 0{]}': 0.009, \\ '{[}3 0 0{]}': 0.0033, '{[}0 0 1{]}': 0.0029, '{[}2 1 0{]}': 0.0019, '{[}1 0 1{]}': 0.0012, '{[}0 2 0{]}': 0.0011, \\ '{[}1 2 0{]}': 0.0007, '{[}0 1 1{]}': 0.0005, '{[}3 1 0{]}': 0.0004, '{[}1 1 1{]}': 0.0003, '{[}2 0 1{]}': 0.0003, \\ '{[}4 0 0{]}': 0.0003, '{[}2 2 0{]}': 0.0003, '{[}0 3 0{]}': 0.0001, '{[}1 2 1{]}': 0.0001, '{[}5 0 0{]}': 0.0001\}\\ Others: 0.0004\end{tabular}}} \\ \hline
\end{tabular}
}
\end{center}
\label{osd_pattern_distribution_128}
\end{table}

At high SNR=3.0dB of LDPC (128,64) code, empirical evidence suggests that the soft information in the $T$-th NMS iteration is more reliable than in the $0$-th iteration. Therefore, the former is chosen as the benchmark. Table~\ref{osd_pattern_distribution_128} enumerates the potential dominant order patterns with and without the aid of DIA, where the ratios are obtained by dividing each counter value by the total counts. It is observed that more decoding failures are categorized into fewer dominant order patterns with the assistance of DIA. Specifically, the total ratio of 0.0086 decoding failures falls outside the scope of OSD with $l_{pt}=20$ without DIA. In contrast, the risk reduces to 0.0004 with DIA involvement. For a very small $l_{pt}=2$, the presence of DIA elevates the cumulative ratio sum of '[0,0,0]' and '[1,0,0]' order patterns from 0.8299 to 0.9362, illustrating its concentration effect. A similar observation holds true for other varied $l_{pt}$, whose trend aligns with the CDF illustration of Fig.~\ref{order_distribution_128_2dot2_3dot2}.

On the other hand, as shown in Table~\ref{osd_pattern_distribution_128}, even with the absence of DIA, the decoding path is nearly identical to the case with DIA aid, up to a few swaps in the adjacent range, showcasing its robustness to varied reliable metrics.

Moreover, from some perspectives, the conventional order-$p$ OSD adheres to a special decoding path in our context. For instance, the conventional order-1 OSD follows the unaltered list of order patterns of '[0,0,0]', '[1,0,0]', '[0,1,0]', and '[0,0,1]', with equivalent performance and complexity as our adaptive OSD with $l_{pt}=4$. However, with the increase of order $p$, to avoid the rigidity of conventional OSD, the proposed decoding path-guided OSD can readily customize many new trade-offs with the compromised decoding performance between conventional order-$p$ and order-$(p+1)$ OSDs, at the cost of substantially decreased complexity.

In practice, it is preferable to assign a higher value to $\lambda_{1m}$ and a lower value to $\lambda_{3m}$ to strike a balance between performance and complexity. Moreover, with a fixed decoding path, imposing additional constraints on these $\lambda_{im}$ values can help skip unexpected order patterns along the decoding path, thus managing complexity more effectively. For example, setting $[\lambda_{1m},\lambda_{2m},\lambda_{3m}]$ = $[3,2,1]$, with or without DIA, the constraint $\lambda_m = \sum_{i=1}^{3}\lambda_{im} = 3$ will filter out order patterns like '[3,1,0]', '[4,0,0]', and '[2,2,0]' from the decoding path, as shown in Table~\ref{osd_pattern_distribution_128}. In this scenario, the adopted decoding path for fulfilling the decoding task is essentially a trimmed version of the nominal decoding path.

Further simulations demonstrate the quasi-universality of the chosen decoding path across a wide range of SNR points of interest. Additionally, to streamline the process of acquiring the decoding path, a path obtained for one specific code may perform well for other codes, provided rational partitions are maintained.

Equipped with a variable $l_{pt}$ decoding path and a dynamic mechanism of adding constraints to facilitate the selection of order patterns, the adaptive OSD maintains the decoding performance of conventional OSD and preserves its parallelizable processing capability, while minimizing the drawback of high complexity. Furthermore, the proposed OSD variant enables accurate estimation of the required computing resources to achieve a desired decoding performance, and vice versa.

\subsubsection{Adaptive OSD Algorithm}
The pseudo-code for the adaptive OSD algorithm is provided in Alg.~\ref{alg::renewed_osd}, with the modified sections highlighted in blue, as compared to the conventional order-$p$ OSD outlined in Section \ref{preliminary}.

As mentioned in step 2 of Alg.~\ref{alg::renewed_osd}, it's important to note that swapping resulting from GE requires a reordering of the MRB bits in ascending magnitude. This aims to concentrate error-prone bits in the leftmost segment of MRB, which in turn triggers corresponding column swapping in $\mathbf{Q_2}$. Consequently, the MRB bits undergo an additional rearrangement compared to the LRB bits. Steps 4, 5, and 7 focus on identifying suitable order patterns for generating all associated TEPs. Additionally, an auxiliary criterion, elaborated in the following subsection, is introduced in step 8 to reduce computations required by criterion ~\ref{argmin_dis}. Notably, the 'for' loop spanning steps 4 to 13, along with step 14 for finding the optimal estimate, can be efficiently implemented using vector computing to minimize response latency.
\begin{algorithm}[ht]
\caption{Decoding path-guided OSD Algorithm}
\label{alg::renewed_osd}
\begin{algorithmic}[1]
\color{blue}
\Require
\color{black}
Received sequence $\mathbf{y}$,
Parity check matrix $\mathbf{H}$,
\color{blue}
Reliability metric $\tilde{\mathbf{y}} = [\tilde{y}_i]_1^N$ from DIA output,
Predetermined decoding path of length $l_{pt}$,
and the constraints on 
$[\lambda_{1m},\lambda_{2m},\lambda_{3m}]$.
\color{black}
\Ensure
Optimal codeword estimate $\hat{\mathbf{c}}$
\State
Sort $[\tilde{y}_i]_1^N$ in ascending order of magnitude to obtain $[\tilde{y}_i^{(1)}]_1^N$. Simultaneously, $\mathbf{H^{(0)}}$=$\mathbf{H}$ is transformed into $\mathbf{H}^{(1)}$.
\State
Reduce $\mathbf{H}^{(1)}$ to its systematic form $\mathbf{H}^{(2)} = \mathbf{\left [I:Q_2\right ]}$. The requested column swapping and its resultant ordering of MRB bits lead to the rearrangement of $[\tilde{y}_i^{(1)}]_1^N$ bits into $[\tilde{y}_i^{(2)}]_1^N$.
\State
Obtain the hard decision $\mathbf{c}_a$ of $[\tilde{y}_i^{(2)}]_{M+1}^N$ as the anchoring point.
\color{blue}
\For {$l \in \{1,2,\ldots,l_{pt}\}$}
\If{the $l$-th order pattern satisfies the constraints on $[\lambda_{1m},\lambda_{2m},\lambda_{3m}]$}
\color{black}
\Repeat:
\State
Generate one TEP $\mathbf{e}_j$ \color{blue} of the $l$-th order pattern\color{black}, and $\mathbf{e}_j \oplus \mathbf{c}_a \Rightarrow {\overline{\mathbf{c}}_{2j}}$, $\overline{\mathbf{c}}_{1j} = \overline{\mathbf{c}}_{2j}\mathbf{Q_2^\top}$
\color{blue}
\If{criterion \ref{aux_qualify} met}\color{red}  \hspace{1cm} (optional)
\color{black}
\State
Append the candidate $\overline{\mathbf{c}}_{j} = [\overline{\mathbf{c}}_{1j}:\overline{\mathbf{c}}_{2j}]$ into the codeword candidates list.
\color{blue}
\EndIf
\color{black}
\Until{traversing ends}
\color{blue}
\EndIf
\EndFor
\color{black}
\State
Select the optimal $\tilde{\mathbf{c}}$ in the codeword candidates list akin to criterion~\ref{argmin_dis}, then reverse all the involved bit swaps for $\tilde{\mathbf{c}}$ to obtain $\hat{\mathbf{c}}$.
\State
\Return {the codeword estimate $\hat{\mathbf{c}}$ for $\mathbf{y}$}.
\end{algorithmic}
\end{algorithm}

\subsection{Auxiliary Criterion for Downsizing Candidates}

Considering that column swapping induced by GE operations removes $n_{sw}$ bits of high reliability from the MRB, the hard decisions made on these bits will result in few errors. It's intuitive that when an encoded codeword candidate represents the ground truth, the discrepancy between the hard decisions on the positions of these expelled bits and the encoded codeword candidate in the same positions will be small in terms of the number of differing bits. To address this, we can identify $\psi_2(\cdot)$ bit positions with the highest reliability in the LRB and set a threshold $\psi_1(\cdot)$ for the discrepancy between the encoded codeword estimate and the hard decision on these positions, where $\psi_{1,2}(\cdot)$ denote the respective functions of $\lambda_m$ and $n_{sw}$. All codeword estimates exceeding the threshold will be discarded immediately. Therefore, the auxiliary assessment, termed auxiliary criterion ~\ref{aux_qualify}, is formulated as follows:

\begin{equation}
\label{aux_qualify}
\sum_{l=1}^{\psi_2(\cdot)} \mathbf{1}\left ( c^{(l)}  \oplus  \overline{c}_{1j}^{(l)} \right ) \leq \psi_1(\cdot)
\end{equation}

Here, $c^{(l)}$ refers to the hard decision of bits on the $\psi_2(\cdot)$ bit positions, and $\overline{\mathbf{c}}_{1j}$ indicates the bits in the same positions for the encoded codeword candidates. It's worth noting that the selected bit positions are often not adjacent to each other due to the lack of sorting in the LRB  after GE operation. These hyperparameters need to be determined empirically. For simplicity, $\psi_1(\cdot)=\lambda_m$ and $\psi_2(\cdot)=3\lambda_m$ are suggested in the following simulations for the LDPC (128,64) code. The effectiveness of this auxiliary criterion will be further discussed in the complexity analysis subsection.

%% file: sections/simulations.tex
\section{Experimental Results and Complexity Analysis}
\label{simulations}
\subsection{Parameter Settings of Codes}
Five linear block codes will be discussed, including three LDPC codes of short block lengths: (64,32) code, TU KL (96,48) code, and CCSDS (128,64) code, all with a rate of $0.5$. Particularly, the CCSDS (128,64) code is of particular interest as it has been extensively discussed in related literature, facilitating comparisons between the proposed design and existing schemes in terms of decoding performance and complexity. Additionally, a longer WiMAX (802.16) high rate $0.83$ (1056,880) code and a classical linear BCH (63,36) code with a rate of $0.57$ will be examined to ascertain the boundaries and limits of the proposed scheme. The associated parity check matrices $\mathbf{H}$s of the codes in '.alist' format and the ML performance of each code can be found in the code database or links provided in the literature \cite{helmling19, Rosseel2022}.

Table \ref{para_setting} presents parameter settings for the NMS-DIA-OSD decoding framework, where empirical evaluations are denoted in black and trainable parameters in red. The maximum number of iterations $T$ significantly affects both the NMS decoding performance and the effectiveness of the DIA model. A larger $T$ for NMS generally yields better FER but may result in longer latency in worst cases. Moreover, a long fluctuating trajectory of decoding failures may hinder the DIA from forging informative metrics, adversely impacting the adaptive OSD. Therefore, a rational evaluation of $T$ is chosen empirically to maintain the trade-off between performance and complexity. The choice of training SNR is typically located at the SNR point where the resulting NMS FER is around $10^{-1}$ to generate training samples for the DIA model. It was found that generating blended batches of training samples across a range of SNRs was slightly inferior. Similarly, a trial-and-error strategy is applied to fine-tune other parameters.

Given the well-trained NMS model, DIA model, and available decoding path for the adaptive OSD, all simulations were conducted on the TensorFlow platform, and the related source code will be made available on GitHub at a later time.

\begin{table}[htbp]
\caption{Hyperparameter Settings for All Tested Codes}
\label{para_setting}
\begin{tabular}{|c|c|c|c|c|c|l}
\hline
Codes &
  \begin{tabular}[c]{@{}c@{}}Training \\ SNR\end{tabular}& $T$(NMS) &
  \textcolor{red}{$\alpha$}  &
    \begin{tabular}[c]{@{}c@{}}MRB \\ partition\end{tabular} & $Q$ \\  \hline
(64,32)    & 3.1dB& 8  &  \textcolor{red}{0.97}   &[1/6,1/3,1/2]*$K$&3  \\ \hline
(96,48)    & 2.9dB& 10 & \textcolor{red}{0.91}   & [1/6,1/3,1/2]*$K$&3 \\ \hline
(128,64)   & 2.7dB& 12   & \textcolor{red}{0.78}  & [1/6,1/3,1/2]*$K$&3  \\ \hline
(1056,880)  & 3.3dB& 20 & \textcolor{red}{0.85}  &\begin{tabular}[c]{@{}c@{}}[1/20,1/10,... \\ 17/80,17/40]*$K$\end{tabular} &10 \\ \hline
BCH(63,36)  & 3.2dB& 8  & \textcolor{red}{0.59}  & [1/6,1/3,1/2]*$K$ &3\\ \hline
\end{tabular}
\end{table}
\subsection{Flow Chart of the NMS-DIA-OSD Decoding Framework}
\input{tikz/flowchart_nnms_dia_osd}

As depicted in the block diagram shown in Figure \ref{flow_chart_nms_dia_osd}, NMS decoding operates on received sequences in batch mode. Additionally, the entire decoding process terminates immediately upon meeting the early stopping criterion $\mathbf{H}\hat{\mathbf{c}}^\top$ == $\mathbf{0}$. For the trajectories collected from NMS decoding failures that do not meet the criterion until the final decoding iteration, DIA is employed to combine them and produce enhanced reliability measurements for each bit in the sequence. These measurements are then inputted into the adaptive OSD for postprocessing. Denoting FER1 as the decoding metric for NMS and FER2 for the adaptive OSD, the overall FER of the framework is the product of FER1 and FER2. It's worth noting that while undetectable decoding failures were observed in the simulations of the first two short codes in low SNR regions, their occurrences are at least two orders of magnitude lower than the counts of detectable failures, thus they are safely ignored in the waterfall SNR region of interest.

The proposed decoding flow chart bears similarity to the one proposed in \cite{baldi2016use}. However, it introduces DIA as a new component and replaces the conventional OSD with an adaptive OSD. Both approaches share the advantage of aligning the limited processing capacity of OSD with a limited number of decoding failures that fail parallel NMS decoding. This streamlined architecture preserves the benefits of high throughput and superior decoding performance, among others.

\subsection{Decoding Performance}

To assess the impact of DIA, adaptive OSD, or auxiliary criterion on decoding performance, the following simulations aim to achieve either improved decoding performance within a given complexity budget or to maintain competitive decoding performance with minimal complexity. It's important to note that the performance curves, with their sources indicated in the legend, have been meticulously calibrated to ensure a fair and accurate comparison with the proposed scheme. Consequently, we primarily focus on conducting performance and complexity comparisons for codes that have been previously discussed in the literature.

In the collected references, typical scheme notations include: D$_{10}$-O-2(25) \cite{Rosseel2022}, which represents a framework consisting of 10 BP-RNN blocks combined with order-2 OSD with a total iteration $T$=25; NBP-D(10,4,4) \cite{buchberger20}, which denotes a neural BP of $T$=10 integrated with 4 list-based decimations determined by 4 learned decimations through a neural network; PB-NBP-D$_1$ \cite{buchberger20}, a pruning-based neural BP that customizes the associated $\mathbf{H}$ in each iteration and exploits a set of optimized weights of check nodes; and PB-OSD, a probability-based state-of-the-art decoder discussed in the latter part of Section \ref{simulations} for comparison with the proposed approach.

For a decoding path constrained by $\lambda_m$, there exist $l_{pt}=\sum_{i=0}^{\lambda_m}\mathbf{1}(\sum_{j=1}^{Q}n_j\le i)$ nominally ordered patterns, each in the form $\left [n_1,n_2,\cdots,n_Q\right ]$ where $n_i$ denotes the number of non-zero elements in the corresponding segment of the MRB. In the legend of the following plots, a typical notation N-O(1,4) denotes the combination of NMS and conventional OSD \cite{baldi2016use}, while N-D-O(1,4) represents the concatenation of NMS, DIA, and adaptive OSD, both with $\lambda_m=1$ and $\hat{l}_{pt}=4$, where $\hat{l}_{pt}$ is the surviving length of a trimmed nominal decoding path after applying the related constraints. For instance, with $\lambda_{m}=1$ and $l_{pt}=5$, the nominal decoding path is ordered as $[0,0,0],[1,0,0],[0,1,0],[1,1,0],[0,0,1],\cdots$, and the specific order pattern $[1,1,0]$ will be abandoned due to $\sum\lambda_i>\lambda_m$, leading to the notation (1,4). Additionally, regarding the included TEP(s), the decoding paths noted as (0,1) and (1,4) correspond exactly to the conventional order-0 and order-1 OSD, respectively. As benchmarks, the decoding plots of NMS and BP alone are commonly included, with the number of iterations indicated in parentheses. At each SNR point under test, at least 100 decoding failures are collected to ensure plot stability. 

\input{plots/plot_64_fer}
\input{plots/plot_96_fer}
\input{plots/plot_128_fer}

As shown in Fig.~\ref{fer64}, N-D-O(0,1) lags behind D$_{10}$-O-0(25) \cite{Rosseel2022}, but N-D-O(1,4) rapidly catches up with D$_{10}$-O-1(25), both achieving ML performance for the short code. In comparison, conventional BP with a remarkable $T$ setting or NMS alone still fall far behind other curves.

For the LDPC (96,48) code, as depicted in Fig.~\ref{fer96}, in terms of FER at $10^{-3}$, upgrading ($\lambda_m,\hat{l}_{pt}$) from (1,4) to (2,6) narrows the gap to 0.15dB, and further upgrades to (3,8) yield no marked gain due to approaching ML performance. Hence, N-D-O(2,6) is favored for its substantially lower complexity.

When the block length increases to 128, the required computational resources pose challenges for our existing platform before reaching ML performance. As shown in Fig.~\ref{fer128}, the proposed N-D-O(3,15), while comparable to the SOTA D$_{10}$-O-2(25) \cite{Rosseel2022}, lags behind the ML curve by within 0.25dB. Compared with other methods, it significantly outperforms NBP-D(10,4,4) \cite{buchberger21} by over 0.6dB, while the latter, despite its high complexity, only slightly outperforms the proposed N-D-O(1,4) and falls behind the N-D-O(3,10) by about 0.45dB at FER=$10^{-3}$.

\input{plots/plot_1056_fer}

Improving decoding performance for longer codes typically presents more challenges. Firstly, the adaptive OSD is expected to solicit a larger number of order patterns to achieve the targeted decoding FER, as NMS decoding failures are scattered more widely due to the broader MRB width, resulting in no evident concentration effect. Secondly, the escalating complexity of the conventional order-$2$ OSD, although far from achieving ML performance, makes its implementation impractical given limited computational resources. Nonetheless, a simulation of a longer LDPC $(1058,880)$ code was conducted to assess the scalability of our proposed scheme. At the expense of some performance degradation, fine-tuning of $\lambda_m$ and its related constraints is necessary to avoid computation-intensive order patterns encountered in securing the actual decoding path. Additionally, partitioning the MRB into more segments, such as $Q=10$, was attempted for this code. To ensure fairness for unevenly sized order patterns, the sorting criterion for deciding priorities in the nominal decoding path involved arranging them in descending order based on the hit rate, defined as the number of trappings divided by the number of TEPs included. This differs from the criteria adopted for other short codes. As a result, the inherent increase in complexity did affect the adaptive OSD, but to a much lesser extent. As illustrated in Fig.~\ref{fer1056}, the low-complexity NMS with $T=20$ lags behind BP decoding with $T=40$ by about 0.1dB. However, with the support of DIA and adaptive OSD, the N-D-O combinations significantly surpass the latter, and the performance difference between N-D-Os is within 0.1dB, despite the complexity increasing in the order of magnitude sequentially. Unfortunately, there remains a gap of more than 0.25dB between the closest curve of N-D-O(3,50) and the ML curve. Hence, it suggests that a nuance of FER performance enhancement generally entails an expensive complexity cost for codes of long block lengths.

\input{plots/plot_63_fer}

The traditional BCH $(63,36)$ code is a high-density parity-check code with many short cycles in its code structure. As shown in Fig.~\ref{fer63}, whether with or without DIA, the corresponding N-Os and N-D-Os show no discernible performance difference before reaching ML performance. To trace the cause, it is conjectured that due to the poor performance of NMS, with its FER around 0.5 throughout the tested SNR region, the trajectories of its decoding failures hold no informative clues for the DIA to find a way to improve bit reliability metrics.

\subsection{Ablation Analysis of DIA} 

\input{plots/plot_128_fer_ablation}

For the LDPC $(128,64)$ code, as shown in Fig.~\ref{fer128ablation}, PB-NBP $D_1$ \cite{buchberger20} outperforms both NBP and NMS significantly, but it is slightly surpassed by the combination 'N-O-(2,8)'. Additionally, 'NBP-D(10,4,4)' \cite{buchberger21} demonstrates comparable performance to N-O-(3,12) in terms of FER. It's noteworthy that with the presence of DIA, N-D-O(2,8) exhibits at least a 0.4dB gain over N-O(2,8), and an additional 0.1dB gain is achieved when upgrading N-D-O(2,8) to N-D-O(3,10). This indicates that the improved bit reliability metric has the potential to significantly enhance the OSD performance.

\subsection{Complexity Analysis}

From a certain perspective, the decoding process of the conventional order-$p$ OSD can be seen as following a predetermined path consisting of a comprehensive list of order patterns, each with a Hamming weight less than or equal to $p$. In contrast, the proposed adaptive OSD determines the precedence of each order pattern based on the occurrences of authentic error patterns associated with it. By partitioning the MRB into $Q$ segments, the order pattern serves as a smaller building unit, allowing for customization to decide whether it should be included in the actual decoding path. This typically results in an incomplete list of order patterns. Additionally, DIA demonstrates a concentration effect for the adaptive OSD by categorizing more error patterns underlying the decoding failures into its dominant order patterns.

Naively, when extending $Q=K$ for the code ($N,K$), or equivalently, when each order pattern corresponds exactly to one error pattern, the granularity reaches its limit. However, this approach is only applicable for short codes, as a large number of samples is required in the query phase to ensure the statistical stability of the occurred error patterns.

\subsubsection{Effect of Auxiliary Criterion}

When the survived order patterns generate TEPs to populate the list of codeword candidates, the newly introduced auxiliary criterion will threshold the potential TEPs, leading to the abandonment of some candidates from further arithmetic operations. This is done in the hope of alleviating complexity demands at the cost of minimal performance loss. Figure~\ref{fer128cutsize} illustrates the impact of the auxiliary criterion on decoding performance for the (128,64) code, where the decoding schemes suffixed with 'S' indicate the application of the auxiliary criterion. It can be observed that the performance variation is negligible for the adaptive OSD in both cases, with or without the criterion. However, as shown in Table~\ref{tab:cutsize-table}, the introduction of the auxiliary criterion can nearly halve the list size of codeword candidates after proper evaluation of $\psi_{1,2}$. Similar observations hold for other codes as well. As a benchmark, despite its simplicity, PB-NBP-$D_1$ \cite{buchberger20} experiences severe performance loss compared to its counterparts.

\input{plots/plot_128_fer_cutsize}

\begin{table}[htbp]
\caption{Comparison of the list sizes of codeword candidates for ($\lambda_m,\hat{l}_{pt}$)=(2,10),(3,12) OSD of LDPC (128,64) code with and without the auxiliary criterion.}
\label{tab:cutsize-table}
\resizebox{\columnwidth}{!}{%
\begin{tabular}{|c|c|c|c|c|c|c|c|}
\hline
($\lambda_m,\hat{l}_{pt}$) & \begin{tabular}[c]{@{}c@{}}Auxiliary\\ Criterion?\end{tabular} & 2.0dB & 2.2dB & 2.4dB & 2.6dB & 2.8dB & 3.0dB \\ \hline
\multirow{2}{*}{(2,10)} & No & 2081 & 2081 & 2081 & 2081 & 2081 & 2081 \\ \cline{2-8}
& \cellcolor[HTML]{FFCCC9}Yes & \cellcolor[HTML]{FFCCC9}1395 & \cellcolor[HTML]{FFCCC9}1415 & \cellcolor[HTML]{FFCCC9}1423 & \cellcolor[HTML]{FFCCC9}1439 & \cellcolor[HTML]{FFCCC9}1452 & \cellcolor[HTML]{FFCCC9}1461 \\ \hline
\multirow{2}{*}{(3,12)} & No & 11120 & 11120 & 11120 & 11120 & 11120 & 11120 \\ \cline{2-8}
& \cellcolor[HTML]{FFCCC9}Yes & \cellcolor[HTML]{FFCCC9}6229 & \cellcolor[HTML]{FFCCC9}6280 & \cellcolor[HTML]{FFCCC9}6361 & \cellcolor[HTML]{FFCCC9}6404 & \cellcolor[HTML]{FFCCC9}6497 & \cellcolor[HTML]{FFCCC9}6578 \\ \hline
\end{tabular}%
}
\end{table}

\subsubsection{Constraints on Order Pattern}

In the context of adaptive OSD, imposing more constraints on $\lambda_m$ or $\lambda_{im}$ can effectively filter potential order patterns. Assuming the absence of an auxiliary criterion and given $\lambda_m=\sum_{i=1}^{Q}\lambda_{im}$ with $Q=3$, the list size of codeword candidates (or TEPs) is calculated as follows:

\begin{equation}
\sum_{\lambda_1=0}^{\lambda_{1m}}\sum_{\lambda_2=0}^{\lambda_{2m}}\sum_{\lambda_3=0}^{\lambda_{3m}} \binom{L_\lambda }{\lambda_1} \binom{M_\lambda }{\lambda_2} \binom{H_\lambda }{\lambda_3}
\label{candidate_size_sum}
\end{equation}

where $L_\lambda, M_\lambda, H_\lambda$ are the segment widths of the partitioned MRB. Hence, one combination of $[\lambda_1,\lambda_2,\lambda_3]$ represents a specific order pattern.

A close examination reveals that the order patterns contribute unequally to the sum in \eqref{candidate_size_sum}. For the LDPC (128,64) code, assuming $L_\lambda=10, M_\lambda=20, H_\lambda=34$, and $\lambda_m=3$, then the order pattern $[3,0,0]$ includes 120 TEPs while $[0,1,2]$ includes 11220 TEPs. Hence, the significant size discrepancy among order patterns necessitates additional fine-tuned constraints $\mathbf{\Phi}(\lambda_{1m},\lambda_{2m},\lambda_{3m})$ on the segments' threshold $[\lambda_{1m},\lambda_{2m},\lambda_{3m}]$, such as:

\begin{align*}
\lambda_{3m}\leq 2 \text{ ; } \quad
\lambda_{1m}=\lambda_{2m}=0 \text{ if } \lambda_{3m}=2.
\end{align*}

It was found that appropriate $\mathbf{\Phi}(\lambda_{1m},\lambda_{2m},\lambda_{3m})$ can result in a substantial reduction in the number of TEPs requested.

\subsection{Complexity comparison with existing LDPC decoders}

For the complexity analysis of various decoders, we choose to overlook the associated binary operations in re-encoding, syndrome checking, and GE involved in OSD, etc., reasonably. Firstly, these binary operations are hardware-friendly. Secondly, the real arithmetic operation involved is comparable with binary operation for each decoder in terms of their occurrences, thus dominating the computational complexity.

Generally, for a hybrid decoding framework, the average complexity $C_{av}$ can be approximated as follows \cite{baldi2016use}:
\begin{equation}
    C_{av} = C_{NMS} + \tau C_{OSD}
    \label{weighted_complexity}
\end{equation} 
where $C_{NMS}$ and $C_{OSD}$ represent the complexity of the respective decoders in the framework. The claim remains valid in that $\tau$ is nearly equal to the FER of NMS for the codes in our scenario, considering the undetectable decoding failures are rare thus safe to neglect them, while the adaptive OSD plays the role of postprocessing upon NMS decoding failures only.

Suppose the received sequences of an LDPC $(N,K)$ code with $T$ iterations of decoding, given the average column and row weights of its $\mathbf{H}$ as $w_c$ and $w_r$ respectively. The respective complexity constituents for some typical decoders and the proposed are summarized in Table~\ref{tab:complexity-table}, where the implementation complexity expressions are from the sourced references. 

\begin{table}[htbp]
\caption{The worst-case complexity comparison of various decoding schemes per sequence 
 of LDPC $(N,K)=(128,64)$ code with $(w_c,w_r)=(4,8)$, with the counts of arithmetic operations related or trainable parameters exposed in the sourced references.}
\label{tab:complexity-table}
\resizebox{\columnwidth}{!}{%
\begin{tabular}{|c|c|ccc|c|c|}
\hline
Decoders &
  Phases &
  \multicolumn{2}{c|}{$\#$ of  Additions/Comparisons} &
  $\#$ of Multiplications &
  $\#$ of Divisions &
  \# of trainables \\ \hline
{\color[HTML]{3531FF} } &
  {\color[HTML]{3531FF} \textbf{NMS}} &
  \multicolumn{2}{c|}{{\color[HTML]{3531FF} \textbf{\begin{tabular}[c]{@{}c@{}}$NT(3w_c+2)+$\\
  $(N-K)T(2w_r-3)$\\ $\approx 31.5$K\end{tabular}}}} &
  {\color[HTML]{3531FF} \textbf{\begin{tabular}[c]{@{}c@{}}$(N-K)T$\\ $\approx 0.8$K\end{tabular}}} &
  {\color[HTML]{3531FF} \textbf{0}} &
  {\color[HTML]{3531FF} \textbf{1}} \\ \cline{2-7} 
{\color[HTML]{3531FF} } &
  {\color[HTML]{3531FF} \textbf{AID}} &
  \multicolumn{3}{c|}{{\color[HTML]{3531FF} \textbf{335K \textit{FLOPs}}}} &
  {\color[HTML]{3531FF} \textbf{0}} &
  {\color[HTML]{3531FF} \textbf{159}} \\ \cline{2-7} 
\multirow{-3}{*}{{\color[HTML]{3531FF} \textbf{N-A-O(3,12)-S}}} &
  {\color[HTML]{3531FF} \textbf{OSD}} &
  \multicolumn{2}{c|}{{\color[HTML]{3531FF} \textbf{201K$\sim$214K}}} &
  {\color[HTML]{3531FF} \textbf{0}} &
  {\color[HTML]{3531FF} \textbf{0}} &
  {\color[HTML]{3531FF} \textbf{0}} \\ \hline
BP($T$=12)\cite{Fossorier2001} &
  - &
  \multicolumn{2}{c|}{\begin{tabular}[c]{@{}c@{}}$NT(3N-3K+1)$\\ $\approx 296.4$K\end{tabular}} &
  \begin{tabular}[c]{@{}c@{}}$NT(11N-11K-9)$\\ $\approx 1.068$M\end{tabular} &
  \begin{tabular}[c]{@{}c@{}}$NT(N-K+1)$\\ $\approx 99.8$K\end{tabular} &
  0 \\ \hline
NBP($T$=12)\cite{nachmani16} &
  - &
  \multicolumn{2}{c|}{\begin{tabular}[c]{@{}c@{}}$NT(3N-3K+1)$\\ $\approx 296.4$K\end{tabular}} &
  \begin{tabular}[c]{@{}c@{}}$NT(11N-11K-9+2w_c)$\\ $\approx 1.08$M\end{tabular} &
  \begin{tabular}[c]{@{}c@{}}$NT(N-K+1)$\\ $\approx 99.8$K\end{tabular} &
  12288 \\ \hline
PB-NBP-$D_1$($T$=25)\cite{buchberger20} &
  - &
  \multicolumn{2}{c|}{\begin{tabular}[c]{@{}c@{}}$2NT(3N-3K+1)$\\ $\approx 1.23$M\end{tabular}} &
  \begin{tabular}[c]{@{}c@{}}$2NT(11N-11K-9)$\\ $\approx 26.7$M\end{tabular} &
  \begin{tabular}[c]{@{}c@{}}$2NT(N-K+1)$\\ $\approx 2.49$M\end{tabular} &
  28416 \\ \hline
NBP-$D$(10,4,4)($T$=50)\cite{buchberger21} &
  - &
  \multicolumn{2}{c|}{\begin{tabular}[c]{@{}c@{}}$18.2NT(3N-3K+1)$\\ $\approx 11.25$M\end{tabular}} &
  \begin{tabular}[c]{@{}c@{}}$18.2NT(11N-11K-9)$\\ $\approx 242.97$M\end{tabular} &
  \begin{tabular}[c]{@{}c@{}}$18.2NT(N-K+1)$\\ $\approx 22.71$M\end{tabular} &
  1553 \\ \hline
 &
  $D_{10}$ &
  \multicolumn{2}{c|}{\begin{tabular}[c]{@{}c@{}}$10NT(3N-3K+1)$\\ $\approx 6.18$M\end{tabular}} &
  \begin{tabular}[c]{@{}c@{}}$10NT(11N-11K-9)$\\ $\approx 133.5$M\end{tabular} &
  \begin{tabular}[c]{@{}c@{}}$10NT(N-K+1)$\\ $\approx 12.48$M\end{tabular} &
  10240 \\ \cline{2-7} 
\multirow{-2}{*}{$D_{10}$-OSD-2($T$=25)\cite{Rosseel2022}} &
  OSD-2 &
  \multicolumn{2}{c|}{69.02K} &
  0 &
  0 &
  0 \\ \hline
\end{tabular}
}
\end{table}

\subsection{Complexity Comparison with Existing LDPC Decoders}

Regarding the worst-case complexity, we can sum up the involved arithmetic operations or FLOPs (defined as floating point operations) by row to make rough comparisons among them, except that the $D_{10}$-OSD-2(T=25) and the proposed framework are the sum of two and three sub-rows respectively. Evidently, the proposed framework is comparable to BP or NBP in terms of the count of arithmetic operations. Its advantage is extended when compared with the remaining three decoders in the table which are overwhelmed by multiplication operations.

When comparing the average complexity, the average iteration of all decoders is highly dependent on the SNR of interest. Thus, the data in Table.~\ref{tab:complexity-table} should be revised by substituting the averaged $T$ in place. Specifically, the SOTA $D_{10}$-OSD-2(T=25) is the weighted sum of two sub-rows depending on its parallel or serial working mode, and the proposed is the weighted sum of three sub-rows where the extra DIA complexity is merged into $C_{OSD}$ as shown in \eqref{weighted_complexity}. Hence, the complexity of the NMS part of the proposed will dominate the total complexity considering that $\tau$ in \eqref{weighted_complexity} is typically tiny, thus more advantageous than its counterparts.

Secondly, considering the number of trainable parameters is an important index to evaluate model complexity, our DIA employs a compact CNN model with 159 parameters, while NBP-$D$(10,4,4) has 1553 parameters, and other NN decoders, including the SOTA $D_{10}$-OSD-2, require more than 10K parameters. Apparently, the proposed DIA is the simplest one among the others, which implies the least hardware complexity.

Lastly, with respect to data throughput, the proposed framework benefits from its streamlined architecture and only slightly lags behind the fully parallelizable BP or its neural network adaptations due to the included adaptive OSD in the serial processing mode. Despite this, we argue the adaptive OSD distinguishes itself from other OSD variants for its characteristic of parallel TEPs processing.

\subsection{Elaborated Comparison with the SOTA PB-OSD}

For the extended family of BCH (eBCH) codes, with solid theoretical support \cite{yue2021revisit}, the PB-OSD scheme stands out as one of the cutting-edge $\mathbf{G}$-oriented decoders due to its minimal consumption of TEPs, which are treated as dominating total complexity \cite{yue2021probability}. Therefore, it intrigues us to substitute PB-OSD for adaptive OSD in the proposed framework to explore its extended effectiveness for LDPC codes. To this end, we employ PB-OSD alone to decode the same collection of NMS decoding failures for the purpose of validation.

From a practical perspective, terminating OSD decoding at the earliest occurrence of the most promising TEP is desirable. Premature termination causes performance loss, whereas delayed termination leads to wasted resources and increased complexity. In practice, the majority of designs, as surveyed in Section \ref{intro_sec}, are based on customizing TEP scheduling on the fly for each received sequence to reduce the number of TEPs. As one of its representatives, PB-OSD holds two critical thresholds $P_t^{suc}$ and $P_t^{pro}$ to discern whether the current TEP will be the most probable error pattern underlying the decoding failure among the remaining TEPs. For fair comparison, we strictly adhere to its description in \cite{yue2021probability} to implement PB-OSD in Python.
\input{plots/plot_128_PB_OSD}
\input{plots/pb_osd_average_teps}

For the experimental results of LDPC (128,64) code, Figure~\ref{fer128_pb_osd} illustrates the FER performance for the schemes under varied conditions, while the average number of TEPs $n_{at}$ is shown in Figure~\ref{fer128_pb_average_tep}. The legend entry D-O(3,8) (O(2,8)) denotes the ($\lambda_m,\hat{l}_{pt}$)=(2,8) adaptive OSD with (without) DIA aid. PB$_0$ denotes the original evaluation for both thresholds of PB-OSD in \cite{yue2021probability}, while PB$_1$ stands for the adapted evaluation of lowering $P_t^{pro}$ to one-tenth of its original value, and PB$_2$ for updating $P_t^{suc}$ to $\max(P_t^{suc},0.9)$ in addition to the alteration for PB$_1$.

Several observations can be drawn from Figs.~\ref{fer128_pb_osd} and \ref{fer128_pb_average_tep}. Firstly, the introduction of DIA significantly improves the proposed framework compared to the case of its absence. For instance, D-O(2,8) outperforms O(3,12) in terms of both FER and $n_{at}$. Secondly, the evaluation of two thresholds impacts the respective FERs of PB-OSD considerably. For example, PB$_2$ is favored over PB$_1$ for achieving better FER at the cost of a slight increase in complexity, thus necessitating further fine-tuning of the thresholds. Thirdly, while PB$_2$ outperforms N-O(2,8), it lags behind N-O(3,12)-S in most SNR regions. Additionally, the family of PB-OSDs as a whole demonstrates a slower declining FER trend than that of adaptive OSDs. Lastly, as illustrated in Fig.~\ref{fer128_pb_average_tep}, PB-OSDs evidently require the fewest $n_{at}$ TEPs when compared with adaptive OSDs, especially in high SNR regions.

On the other hand, complexity is not limited to the counting of TEPs; it is more natural to consider the FLOPs involved in all arithmetic operations. In the case of PB-OSD, it involves many complex operations, such as the product of a long sequence of real numbers and Q-function calculations. Roughly, prior to comparing with threshold $P_t^{pro}$, acquiring the probability of each TEP requires Q-function approximation and at least $N$ FLOPs to prepare for the arguments of the approximation function. Deciding whether to update the minimum weighted Hamming weight consumes about $\frac{1}{3}N$ FLOPs. Then, the overhead of maintaining the underlying heap data structure to generate the next TEP includes two comparisons, while the complexity associated with probability computing of the promising TEPs before $P_t^{suc}$ is complex but rarely occurs, thus negligible. As a result, in the interested range of 2.2dB to 3.2dB, with the loose complexity lower bound $(\frac{4}{3}N+2) \cdot n_{at}$, the total complexity of PB$_2$ ranges from 41K to 189K FLOPs, where $n_{at}$ varies from 238 to 1092. In comparison, for adaptive OSD, benefiting from the effect of auxiliary criterion by reducing the list size of TEPs as shown in Table~\ref{tab:cutsize-table}, the total complexity of D-O(3,12)-S is obtained by summing up the complexity of DIA and that of calling \eqref{argmin_dis} as shown in Table~\ref{tab:complexity-table}, ranging from 536K to 549K FLOPs. Hence, we conclude that the complexity gap between PB-OSD and adaptive OSD is actually narrower than what is conveyed by $n_{at}$ in Fig.~\ref{fer128_pb_average_tep}.

For the worst-case complexity comparison, depending on the order-$p$ setting, PB-OSD exhibits a large variance between its $n_{at}$ and the number of all possible TEPs under the Hamming weight constraint $p$. In contrast, since adaptive OSD follows a predetermined sequence of TEPs of fixed length for any decoding failures, its worst-case complexity shows no variance compared to its $n_{at}$, making it more foreseeable when estimating the maximum decoding latency is necessary.

Lastly, it is worth highlighting that adaptive OSD possesses the characteristic of independence from noise variance, while PB-OSD heavily relies on accurate noise variance to calculate various probabilities. Particularly, in the context of the hybrid framework, the preceding NMS is also invariant to noise variance, making it more consistent when followed by adaptive OSD rather than PB-OSD. Another attractive characteristic of adaptive OSD is its high data throughput due to parallel processing of TEPs, commonly faster than PB-OSDs by more than two orders of magnitude in terms of time complexity in experiments.

Unfortunately, attempts to merge DIA into PB-OSD procedures have not yielded satisfying results. It is conjectured that the forged metric of DIA has disrupted the proper assumptions underlying PB-OSD's operation. Regardless, considering DIA's adeptness at prioritizing competing TEPs statistically, it may be leveraged to adapt existing SOTA OSD variants into at least partially parallel processing modes to reduce decoding latency.

%% file: tikz/flowchart_nnms_dia_osd.tex
\begin{figure}[htbp]
	\centering	
 \tikzstyle{startstop} = [rectangle, rounded corners, minimum width = 2cm, minimum height=1cm,text centered, draw = black, fill = red!40]
	\tikzstyle{io} = [trapezium, trapezium left angle=70, trapezium right angle=110, minimum width=2cm, minimum height=1cm, text centered, draw=black, fill = blue!40]
	\tikzstyle{process} = [rectangle, minimum width=3cm, minimum height=1cm, text centered, draw=black, fill = yellow!50]
	\tikzstyle{decision} = [diamond, aspect = 3, text centered, draw=black, fill = green!30]
	\tikzstyle{arrow} = [->,>=stealth]
	\tikzstyle{bag} = [align=center] 
	\begin{tikzpicture}[node distance=1.4cm,font=\fontsize{9}{9}\selectfont]
		\node (start) [startstop] {Received sequence as input};
		\node (pro1) [process, below of=start,bag] {One NMS iteration \\
yields a tentative $\hat{\mathbf{c}} $};
  	\node (dec1) [decision, 
        below of=pro1,bag] {$\mathbf{H}\hat {\mathbf{c}}^T$ == $\mathbf{0}$ ?};
  	\node (dec2) [decision, 
        below of=dec1,bag] {Max iterations?};
  \node (pro2) [process, below of=dec2] {DIA processing};
		\node (pro3) [process, below of=pro2,bag] {Adaptive OSD to \\query the optimal $\hat {\mathbf{c}}$};
		\node (stop) [startstop, below of=pro3] {Output $\hat{\mathbf{c}}$ };	
		\draw [arrow](start) -- (pro1);
		\draw [arrow](pro1) -- (dec1);
		\draw [arrow](dec1) -- ($(dec1.east) + (0.5,0)$) node[anchor=south] {Yes} |- (stop);
		\draw [arrow](dec1) -- node[anchor=west] {No} (dec2);
		\draw [arrow](dec2) -- ($(dec2.west) + (-0.5,0)$) node[anchor=south] {No} |- (pro1);
		\draw [arrow](dec2) -- node[anchor=west] {Yes} (pro2);  
  	\draw [arrow](pro2) --   
        (pro3);
        \draw [arrow](pro3) --   
        (stop);	
	\end{tikzpicture}
\caption{Decoding of NMS-DIA-OSD Framework}
\label{flow_chart_nms_dia_osd}
\end{figure}
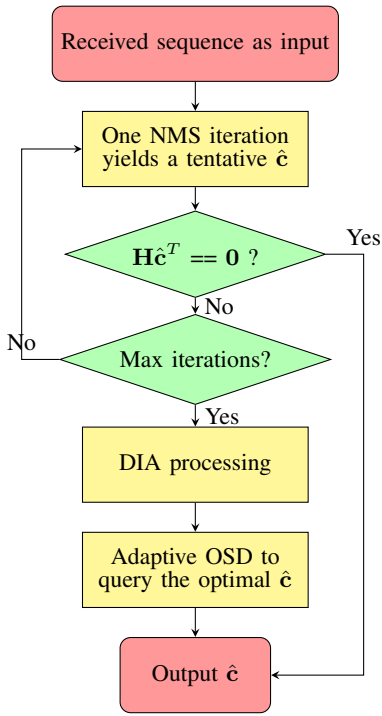

%% file: plots/plot_64_fer.tex
\begin{figure}
 \begin{tikzpicture}
		\begin{semilogyaxis}[
			scale = 0.75,
			xlabel={$E_b/N_0$(dB)},
			ylabel={FER},
			xmin=2.5, xmax=4.1,
			ymin=3e-4, ymax=2e-1,
			xtick={1.0,1.5,2,2.5,...,4.8},
			legend pos = north east,
			ymajorgrids=true,
			xmajorgrids=true,
			grid style=dashed,
			legend style={font=\tiny},
			]
\addplot[
color=violet,
mark=x,
dashed,
very thin
]
coordinates {
(2.6,0.1890)
(2.8,0.1493)
(3.0,0.1139)
(3.2,0.0847)
(3.4,0.0613)
(3.6,0.0409)
(3.8,0.024)
(4.0,0.014)
};	
\addlegendentry{NMS(8)}

\addplot[
color=blue,
mark= +,
dashed,
very thin
]
coordinates {
(3.0,0.053)
(3.25,0.033)
(3.5,0.019)
(3.75,0.01)
(4.0,0.0055)
};	
\addlegendentry{BP(250)\cite{Rosseel2022}}
\addplot[
color=red,
mark=triangle,
very thin
]
coordinates {
(2.6, 0.04521)
(2.7, 0.03713)
(2.8, 0.03021)
(2.9, 0.02558)
(3.0, 0.02239)
(3.1, 0.01775)
(3.2, 0.01282)
(3.3, 0.01094)
(3.4, 0.00896)
(3.5, 0.00677)
(3.6, 0.00561)
(3.7, 0.00419)
(3.8, 0.00317)
};	
\addlegendentry{N-D-O(0,1)}

\addplot[
color=black,
mark=square,
very thin
]
coordinates {
(2.6, 0.01526)
(2.7, 0.0126)
(2.8, 0.01056)
(2.9, 0.00759)
(3.0, 0.00651)
(3.1, 0.00496)
(3.2, 0.00399)
(3.3, 0.00285)
(3.4, 0.00257)
(3.5, 0.00187)
(3.6, 0.00148)
(3.7, 0.0009)
(3.8, 0.00068)
};
\addlegendentry{N-D-O(1,4)}
\addplot[
color=magenta,
mark=halfcircle,
]
coordinates {
  (3.0, 1.2e-02)
  (3.25,6.40e-3)
  (3.5, 3.5e-03)
  (3.75,1.7e-3)
  (4.0, 7.8e-04)
};
\addlegendentry{$D_{10}$-O-0(25)\cite{Rosseel2022}}
\addplot[
color=purple,
mark=*,
very thin
]
coordinates {
  (3.0, 5.7e-03)
  (3.25,3.1e-3)
  (3.5, 1.6e-03)
  (3.75,7.38e-4)
  (4.0, 3.9e-04)
};	
\addlegendentry{$D_{10}$-O-1(25)\cite{Rosseel2022}}
\addplot[
color=cyan,
mark=asterisk,
very thin
]
coordinates {
  (3.0, 5.2e-03)
  (3.25,3.0e-3)
  (3.5, 1.6e-03)
  (3.75,7.35e-4)
  (4.0, 3.6e-04)
};	
\addlegendentry{ML\cite{Rosseel2022}}

		\end{semilogyaxis}
	\end{tikzpicture}
	\caption{FER comparison of decoding schemes for LDPC (64,32) code}
	\label{fer64}
\end{figure}
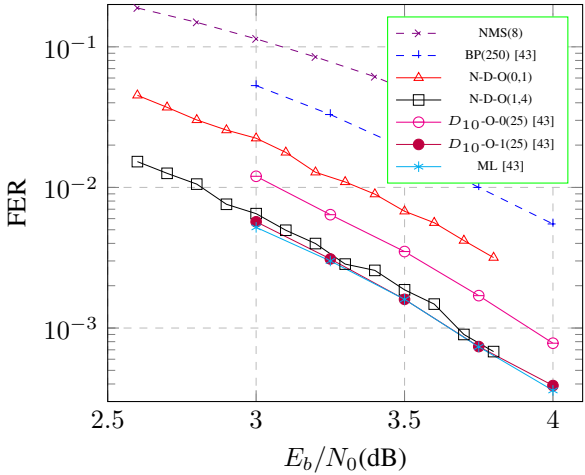

%% file: plots/plot_96_fer.tex
\begin{figure}
 \centering
	\begin{tikzpicture}
		\begin{semilogyaxis}[
			scale = 0.75,
			xlabel={$E_b/N_0$(dB)},
			ylabel={FER},
			xmin=2.4, xmax=3.5,
			ymin=1e-4, ymax=3e-1,
			xtick={1.0,1.5,2,2.5,...,4.8},
			legend pos = north east,
			ymajorgrids=true,
			xmajorgrids=true,
			grid style=dashed,
			legend style={font=\tiny},
			]
\addplot[
color=violet,
mark=x,
dashed,
very thin
]
coordinates {
(2.4,0.2108)
(2.6,0.1589)
(2.8,0.1146)
(3.0,0.0820)
(3.2,0.0529)
(3.4,0.0308)
(3.6,0.015)
};	
\addlegendentry{NMS(10)}

\addplot[
color=blue,
mark= +,
dashed,
very thin
]
coordinates {
(2.5,0.11)
(2.75,0.07)
(3.0,0.04)
(3.5,0.01)
(4.0,2.3e-3)
};	
\addlegendentry{BP(40)\cite{helmling19}}
\addplot[
color=red,
mark=triangle,
very thin
]
coordinates {
(2.4, 0.01808)
(2.5, 0.01541)
(2.6, 0.00898)
(2.7, 0.00816)
(2.8, 0.00563)
(2.9, 0.00461)
(3.0, 0.00324)
(3.1, 0.00269)
(3.2, 0.00182)
(3.3, 0.00111)
(3.4, 0.00094)
};	
\addlegendentry{N-D-O(1,4)}
\addplot[
color=black,
mark=square,
very thin
]
coordinates {
(2.4, 0.01066)
(2.5, 0.00912)
(2.6, 0.00563)
(2.7, 0.00422)
(2.8, 0.00329)
(2.9, 0.00267)
(3.0, 0.00187)
(3.1, 0.00159)
(3.2, 0.00082)
(3.3, 0.00066)
(3.4, 0.00049)
};
\addlegendentry{N-D-O(2,6)}
\addplot[
color=magenta,
mark=halfcircle,
]
coordinates {
(2.4, 0.01008)
(2.5, 0.00857)
(2.6, 0.00508)
(2.7, 0.00362)
(2.8, 0.00281)
(2.9, 0.00238)
(3.0, 0.00156)
(3.1, 0.00113)
(3.2, 0.0007)
(3.3, 0.00054)
(3.4, 0.00039)
};
\addlegendentry{N-D-O(3,8)}
\addplot[
color=cyan,
mark=asterisk,
very thin
]
coordinates {
  (2.0, 2.1e-02)
  (2.25,1.1e-02)
  (2.5,5.6e-3)
  (2.75,3e-3)
  (3.0, 1.5e-03)
  (3.25,6e-4)
  (3.5,2.1e-4)
};	
\addlegendentry{ML\cite{helmling19}}

		\end{semilogyaxis}
	\end{tikzpicture}
	\caption{FER comparison of decoding schemes for LDPC (96,48) code}
	\label{fer96}	
 \end{figure}
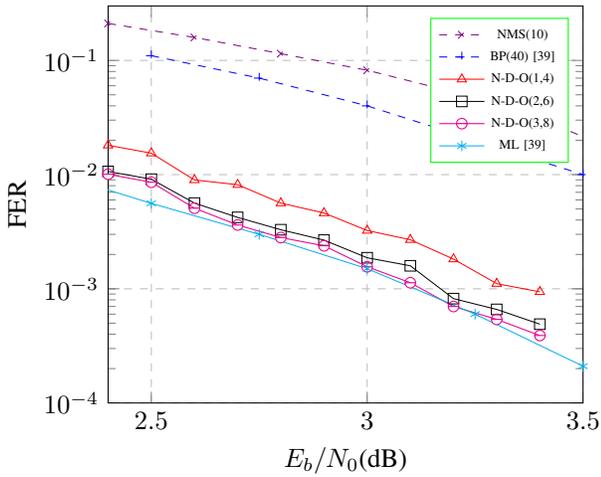

%% file: plots/plot_128_fer.tex
\begin{figure}[htbp]
	\centering
	\begin{tikzpicture}
		\begin{semilogyaxis}[
			scale = 0.75,
			xlabel={$E_b/N_0$(dB)},
			ylabel={FER},
			xmin=1.9, xmax=3.5,
			ymin=1e-4, ymax=1,
			xtick={1.0,1.5,2,2.5,...,4.8},
			legend pos = north east,
			ymajorgrids=true,
			xmajorgrids=true,
			grid style=dashed,
			legend style={font=\tiny},
			]
\addplot[
color=violet,
mark=x,
dashed,
very thin
]
coordinates {
(2.0,0.446)
(2.2,0.360)
(2.4,0.28)
(2.6,0.212)
(2.8,0.150)
(3.0,0.104)
(3.2, 0.068)
};	
\addlegendentry{NMS(12)}
\addplot[
color=blue,
mark= +,
dashed,
very thin
]
coordinates {
(2.0,0.3)
(2.25,0.2)
(2.5,0.13)
(2.75,0.085)
(3.0,0.053)
(3.25,0.027)
(3.5,0.013)
(3.75,0.005)
(4.0,0.002)
};	
\addlegendentry{BP(40)\cite{helmling19}}
\addplot[
color=red,
mark=triangle*,
very thin
]
coordinates {
(2.0, 0.07768)
(2.2, 0.04567)
(2.4, 0.02647)
(2.6, 0.01802)
(2.8, 0.00887)
(3.0, 0.00544)
(3.2, 0.00245) 
};	
\addlegendentry{N-D-O(1,4)}
\addplot[
color=black,
mark=halfcircle,
very thin
]
coordinates {
    (2.0,5e-2)
    (2.5,1.5e-2)
    (3.0,4e-3)
    (3.5,8e-4)
    (4.0,1.2e-4)
};
\addlegendentry{ NBP-D(10,4,4)\cite{buchberger21}}
\addplot[
color=purple,
mark=*,
very thin
]
coordinates {
(2.0, 0.02257)
(2.2, 0.01212)
(2.4, 0.00704)
(2.6, 0.00446)
(2.8, 0.00183)
(3.0, 0.00083)
(3.2, 0.00035)
};
\addlegendentry{N-D-O(3,10)}

\addplot[
color=magenta,
mark=square*,
]
coordinates {
(2.0, 0.01304)
(2.2, 0.00748)
(2.4, 0.00367)
(2.6, 0.00221)
(2.8, 0.00083)
(3.0, 0.00041)
(3.2, 0.00014)
};
\addlegendentry{N-D-O(3,15)}
\addplot[
color=green,
mark=triangle,
very thin
]
coordinates {
  (2.50, 2.2e-03)
  (3.00, 3.0e-04)
  (3.25, 9.0e-05)
  (3.5,2.5e-5)
};	
\addlegendentry{$D_{10}$-OSD-2(25)\cite{Rosseel2022}}
\addplot[
color=cyan,
mark=asterisk,
very thin
]
coordinates {
  (1.00, 1.064e-01)
  (1.50, 3.397e-02)
  (2.00, 8.773e-03)
  (2.50, 1.168e-03)
  (3.00, 1.321e-04)
  (3.50, 1.022e-05)
};	
\addlegendentry{ML\cite{helmling19}}

		\end{semilogyaxis}
	\end{tikzpicture}
	\caption{FER comparison for various decoding schemes of (128,64) code}
	\label{fer128}
\end{figure}
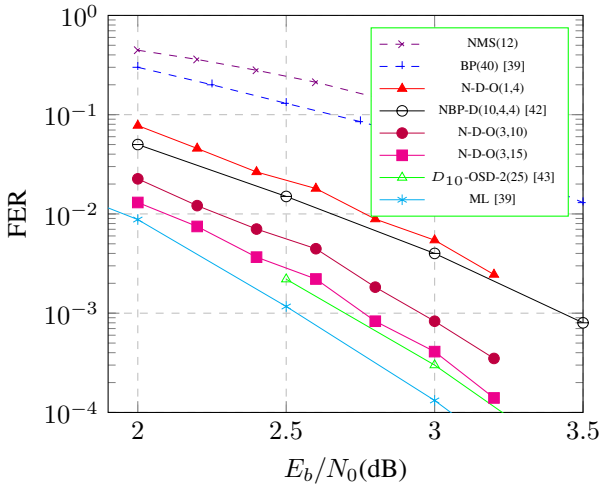	

%% file: plots/plot_1056_fer.tex
\begin{figure}[htbp]
	\centering
	\begin{tikzpicture}
		\begin{semilogyaxis}[
			scale = 0.75,
			xlabel={$E_b/N_0$(dB)},
			ylabel={FER},
			xmin=2.9, xmax=3.8,
			ymin=1e-4, ymax=1,
			xtick={1.0,1.5,2,2.5,...,4.},
			legend pos = south west,
			ymajorgrids=true,
			xmajorgrids=true,
			grid style=dashed,
			legend style={font=\tiny},
			]

\addplot[
color=violet,
mark=x,
dashed,
very thin
]
coordinates {
(3.0,0.4760)
(3.1,0.4040)
(3.2,0.2667)
(3.3,0.1758)
(3.4,0.1094)
(3.5,0.0661)
(3.6,0.0346)
(3.7,0.0169)
(3.8,0.0082)
};
\addlegendentry{NMS(20)}
\addplot[
color=blue,
mark= +,
dashed,
very thin
]
coordinates {
(3.0,0.381875)
(3.1,0.2720192307692308)
(3.2,0.18566176470588236)
(3.3,0.12151041666666666)
(3.4,0.07140116279069768)
(3.5,0.042535714285714284)
(3.6,0.02056451612903226)
(3.7,0.010627104377104377)
(3.8,0.005058061420345489)
};
\addlegendentry{BP(40)\cite{helmling19}}
\addplot[
color=red,
mark=triangle*,
very thin
]
coordinates {
(3.0, 0.18933)
(3.1, 0.13171)
(3.2, 0.0819)
(3.3, 0.0417)
(3.4, 0.02518)
(3.5, 0.01181)
(3.6, 0.00539)
};
\addlegendentry{N-D-O(2,10)}
\addplot[
color=purple,
mark=*,
very thin
]
coordinates {
(3.0, 0.16373)
(3.1, 0.09546)
(3.2, 0.06182)
(3.3, 0.03047)
(3.4, 0.0157)
(3.5, 0.00794)
(3.6, 0.00361)
};
\addlegendentry{N-D-O(3,30)}
\addplot[
color=magenta,
mark=square*,
]
coordinates {
(3.0, 0.12098)
(3.1, 0.08655)
(3.2, 0.044)
(3.3, 0.02518)
(3.4, 0.01226)
(3.5, 0.00554)
(3.6, 0.00248)
};
\addlegendentry{N-D-O(3,50)}

\addplot[
color=cyan,
mark=asterisk,
very thin
]
coordinates {
(3, 4.0e-2)
(3.25, 5.5e-3)
(3.50, 4.0e-04)
(3.75,2.7e-5)
};	
\addlegendentry{ML\cite{helmling19}}
\end{semilogyaxis}
\end{tikzpicture}
\caption{Performance comparison for various decoding schemes of (1056,880) code}
\label{fer1056}
\end{figure}
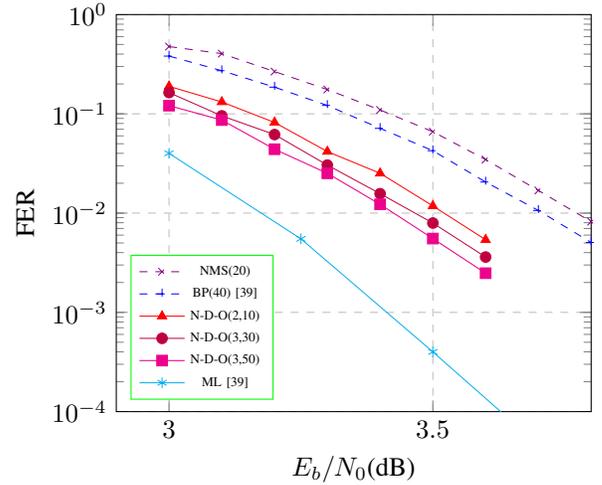	

%% file: plots/plot_63_fer.tex
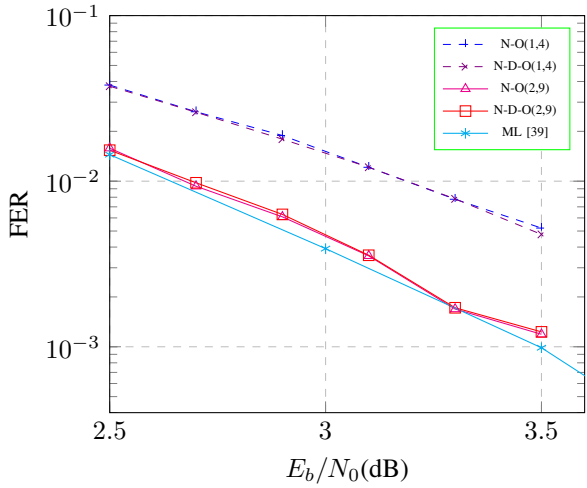
\begin{figure}[htbp]
	\centering
	\begin{tikzpicture}
		\begin{semilogyaxis}[
			scale = 0.75,
			xlabel={$E_b/N_0$(dB)},
			ylabel={FER},
			xmin=2.5, xmax=3.6,
			ymin=4e-4, ymax=1e-1,
			xtick={1.0,1.5,2,2.5,...,4.8},
			legend pos = north east,
			ymajorgrids=true,
			xmajorgrids=true,
			grid style=dashed,
			legend style={font=\tiny},
			]

\addplot[
color=blue,
mark= +,
dashed,
very thin
]
coordinates {
(2.5, 0.03807)
(2.7, 0.02626)
(2.9, 0.01888)
(3.1, 0.01213)
(3.3, 0.00778)
(3.5, 0.00521)
};
\addlegendentry{N-O(1,4)}

\addplot[
color=violet,
mark=x,
dashed,
very thin
]
coordinates {
(2.5, 0.03746)
(2.7, 0.02608)
(2.9, 0.01794)
(3.1, 0.01214)
(3.3, 0.00782)
(3.5, 0.00477)
};
\addlegendentry{N-D-O(1,4)}
\addplot[
color=magenta,
mark=triangle,
]
coordinates {
(2.5, 0.01582)
(2.7, 0.00932)
(2.9, 0.00609)
(3.1, 0.00354)
(3.3, 0.00169)
(3.5, 0.00119)
};
\addlegendentry{N-O(2,9)}
\addplot[
color=red,
mark=square,
]
coordinates {
(2.5, 0.01539)
(2.7, 0.00978)
(2.9, 0.0063)
(3.1, 0.00357)
(3.3, 0.00172)
(3.5, 0.00123)
};
\addlegendentry{N-D-O(2,9)}

\addplot[
color=cyan,
mark=asterisk,
]
coordinates {
  (1.00, 1.739e-01)
  (1.50, 1.018e-01)
  (2.00, 4.482e-02)
  (2.50, 1.447e-02)
  (3.00, 3.916e-03)
  (3.50, 9.864e-04)
  (4.00,1.419e-04)
};	
\addlegendentry{ML\cite{helmling19}}
		\end{semilogyaxis}
	\end{tikzpicture}
	\caption{FER comparison for various decoding schemes of BCH (63,36) code}
	\label{fer63}
\end{figure}	

%% file: plots/plot_128_fer_ablation.tex
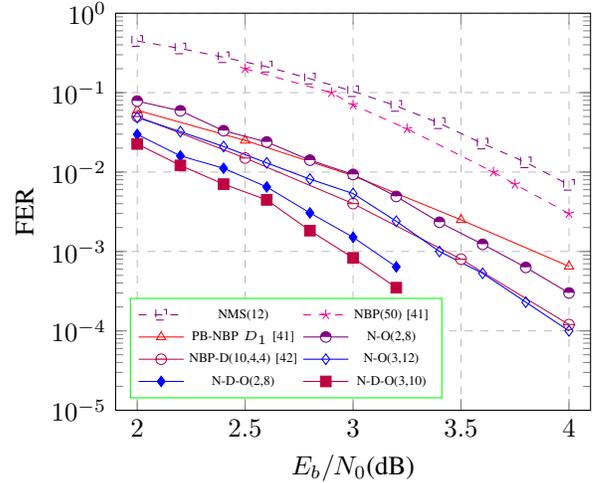
\begin{figure}[htbp]
	\centering
	\begin{tikzpicture}
		\begin{semilogyaxis}[
			scale = 0.75,
			xlabel={$E_b/N_0$(dB)},
			ylabel={FER},
			xmin=1.9, xmax=4.1,
			ymin=1e-5, ymax=1,
			xtick={1.0,1.5,2,2.5,...,4.8},
			legend pos = south west,
			ymajorgrids=true,
			xmajorgrids=true,
			grid style=dashed,
			legend style={font=\tiny,legend columns=2},
			]
\addplot[
color=violet,
mark=square,
very thin,
dashed
]
coordinates {
(2.0, 0.447)
(2.2, 0.36)
(2.4, 0.279)
(2.6, 0.212)
(2.8, 0.15)
(3.0, 0.104)
(3.2, 0.068)
(3.4, 0.04117)
(3.6, 0.02279)
(3.8, 0.01325)
(4.0, 0.0069)
};	
\addlegendentry{NMS(12)}
\addplot[
color=magenta,
mark=star,
very thin,
dashed
]
coordinates {
	(2.5,0.2)
	(2.9,0.1)
	(3.0,0.07)
	(3.25,0.035)
	(3.65,0.01)
	(3.75,0.007)
	(4.0,0.003)
};	
\addlegendentry{NBP(50)\cite{buchberger20}}
\addplot[
color=red,
mark=triangle,
very thin
]
coordinates {
    (2.0,0.06)
    (2.5,0.025)
    (3.0,0.009)
    (3.5,0.0025)
    (4.0,0.00065)
};
\addlegendentry{ PB-NBP $D_1$\cite{buchberger20}}
\addplot[
color=violet,
mark=halfcircle*,
]
coordinates {
(2.0, 0.07823)
(2.2, 0.05899)
(2.4, 0.03315)
(2.6, 0.02391)
(2.8, 0.01411)
(3.0, 0.00932)
(3.2, 0.00493)
(3.4, 0.00233)
(3.6, 0.00123)
(3.8, 0.00063)
(4.0, 0.0003)
};	
\addlegendentry{N-O(2,8)}
\addplot[
color=purple,
mark=halfcircle,
very thin
]
coordinates {
    (2.0,5e-2)
    (2.5,1.5e-2)
    (3.0,4e-3)
    (3.5,8e-4)
    (4.0,1.2e-4)
};
\addlegendentry{ NBP-D(10,4,4)\cite{buchberger21}}

\addplot[
color=blue,
mark=diamond,
]
coordinates {
(2.0, 0.0487)
(2.2, 0.03225)
(2.4, 0.02098)
(2.6, 0.01304)
(2.8, 0.00815)
(3.0, 0.00536)
(3.2, 0.0024)
(3.4, 0.001)
(3.6, 0.00053)
(3.8, 0.00023)
(4.0, 0.00010)
};
\addlegendentry{N-O(3,12)}
\addplot[
color=blue,
mark=diamond*,
very thin
]
coordinates {
(2.0, 0.0298)
(2.2, 0.01597)
(2.4, 0.01116)
(2.6, 0.00649)
(2.8, 0.00306)
(3.0, 0.00151)
(3.2, 0.00064)
};
\addlegendentry{N-D-O(2,8)}
\addplot[
color=purple,
mark=square*,
]
coordinates {
(2.0, 0.02257)
(2.2, 0.01212)
(2.4, 0.00704)
(2.6, 0.00446)
(2.8, 0.00183)
(3.0, 0.00083)
(3.2, 0.00035)
};
\addlegendentry{N-D-O(3,10)}

		\end{semilogyaxis}
	\end{tikzpicture}
	\caption{FER comparison of various decoders of LDPC (128,64) code for the ablation of DIA}
	\label{fer128ablation}
\end{figure}	

%% file: plots/plot_128_fer_cutsize.tex
\begin{figure}[htbp]
	\centering
	\begin{tikzpicture}
		\begin{semilogyaxis}[
			scale = 0.75,
			xlabel={$E_b/N_0$(dB)},
			ylabel={FER},
			xmin=1.9, xmax=3.1,
			ymin=3e-4, ymax=3e-1,
			xtick={1.0,1.5,2,2.5,...,4.8},
			legend pos = north east,
			ymajorgrids=true,
			xmajorgrids=true,
			grid style=dashed,
			legend style={font=\tiny,legend columns=2},
			]
\addplot[
color=red,
mark=triangle,
very thin
]
coordinates {
    (2.0,0.06)
    (2.5,0.025)
    (3.0,0.009)
    (3.5,0.0025)
    (4.0,0.00065)
};
\addlegendentry{ PB-NBP $D_1$\cite{buchberger20}}
\addplot[
color=purple,
mark=halfcircle,
very thin
]
coordinates {
(2.0, 0.02706)
(2.2, 0.01434) 
(2.4, 0.00861)
(2.6, 0.00428)
(2.8, 0.00251) 
(3.0, 0.00122) 
};
\addlegendentry{N-D-O(2,10)-S}
\addplot[
color=black,
mark=*,
very thin
]
coordinates {
(2.0, 0.02591)
(2.2, 0.01426) 
(2.4, 0.00838)
(2.6, 0.0042)
(2.8, 0.00239) 
(3.0, 0.00113) 
};

\addlegendentry{N-D-O(2,10)}
\addplot[
color=blue,
mark=diamond*,
very thin
]
coordinates {
(2.0, 0.01818)
(2.2, 0.00962)
(2.4, 0.00484)
(2.6, 0.00284)
(2.8, 0.00124)
(3.0, 0.00056)
(3.2, 0.00028)
};
\addlegendentry{N-D-O(3,12)-S}
\addplot[
color=magenta,
mark=square*,
]
coordinates {
(2.0, 0.01596)
(2.2, 0.00857)
(2.4, 0.00507)
(2.6, 0.00276)
(2.8, 0.00118)
(3.0, 0.00054)
(3.2, 0.00027)
};
\addlegendentry{N-D-O(3,12)}

		\end{semilogyaxis}
	\end{tikzpicture}
	\caption{FER comparison of various decoders for LDPC (128,64) code with and without auxiliary criterion}
	\label{fer128cutsize}
\end{figure}
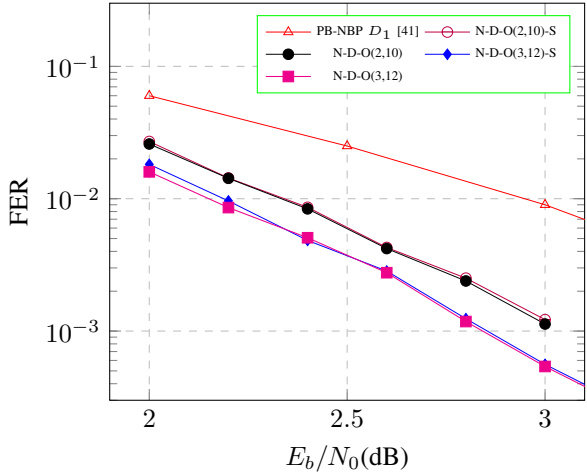	

%% file: plots/plot_128_PB_OSD.tex
\begin{figure}[htbp]
	\centering
	\begin{tikzpicture}
		\begin{semilogyaxis}[
			scale = 0.75,
			xlabel={$E_b/N_0$(dB)},
			ylabel={FER},
			xmin=1.9, xmax=3.3,
			ymin=1e-3, ymax=2e-1,
			xtick={1.0,1.5,2,2.5,...,4.8},
			legend pos = south west,
			ymajorgrids=true,
			xmajorgrids=true,
			grid style=dashed,
			legend style={font=\tiny,legend columns=2},
			]


\addplot[
color=green,
mark=x,
very thin
]
coordinates {
(2.0, 0.175)
(2.2, 0.15385)
(2.4, 0.12882)
(2.6, 0.11278)
(2.8, 0.09409)
(3.0, 0.08957)
(3.2, 0.0725) 
};	
\addlegendentry{O(2,8)}
\addplot[
color=violet,
mark=triangle,
very thin
]
coordinates {
(2.0, 0.06667)
(2.2, 0.05135)
(2.4, 0.04)
(2.6, 0.03061)
(2.8, 0.02041)
(3.0, 0.01449)
(3.2, 0.00935)
};	
\addlegendentry{D-O(2,8)}
\addplot[
color=black,
mark=+,
very thin
]
coordinates {
(2.0, 0.10895)
(2.2, 0.08957)
(2.4, 0.07519)
(2.6, 0.06152)
(2.8, 0.05432)
(3.0, 0.04754)
(3.2, 0.03526)
};
\addlegendentry{ O(3,12)}
\addplot[
color=blue,
mark= halfcircle,
very thin
]
coordinates {
(2.0, 0.04068)
(2.2, 0.02673)
(2.4, 0.01736)
(2.6, 0.01339)
(2.8, 0.00828)
(3.0, 0.00541)
(3.2, 0.00307)
};	
\addlegendentry{D-O(3,12)-S}

\addplot[
color=purple,
mark=*,
very thin
]
coordinates {
(2.2, 0.0463)
(2.4, 0.0431)
(2.6, 0.0412)
(2.8, 0.0379)
(3.0, 0.0336)
(3.2, 0.0304)
};
\addlegendentry{PB$_0$}

\addplot[
color=magenta,
mark=square*,
]
coordinates {
(2.2, 0.0341)
(2.4, 0.0302)
(2.6, 0.027)
(2.8, 0.0243)
(3.0, 0.0224)
(3.2, 0.0204)
};
\addlegendentry{PB$_1$}
\addplot[
color=red,
mark=triangle*,
very thin
]
coordinates {
(2.0,0.0384)
(2.2, 0.0295)
(2.4, 0.023)
(2.6, 0.0183)
(2.8, 0.0152)
(3.0, 0.0125)
(3.2, 0.0095)
};	
\addlegendentry{PB$_2$}

\end{semilogyaxis}
\end{tikzpicture}
\caption{FER performance of LDPC (128,64) code for various schemes under varied conditions.}
\label{fer128_pb_osd}
\end{figure}
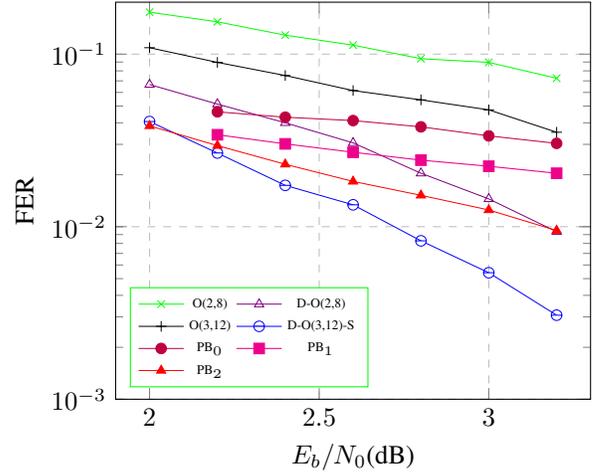	

%% file: plots/pb_osd_average_teps.tex
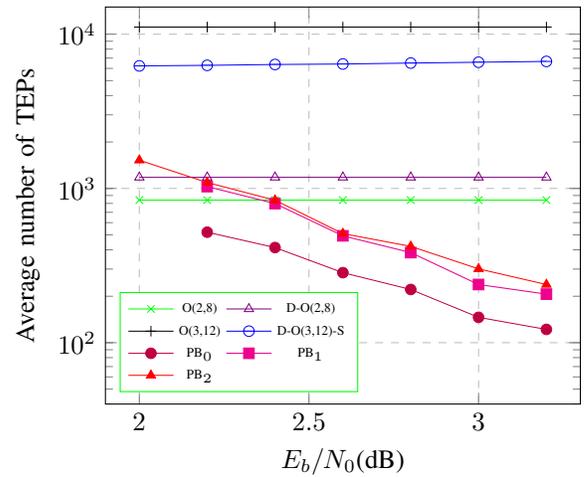
\begin{figure}[htbp]
	\centering
	\begin{tikzpicture}
		\begin{semilogyaxis}[
			scale = 0.75,
			xlabel={$E_b/N_0$(dB)},
			ylabel={Average number of TEPs},
			xmin=1.9, xmax=3.3,
			ymin=4e1, ymax=1.5e4,
			xtick={1.0,1.5,2,2.5,...,4.8},
			legend pos = south west,
			ymajorgrids=true,
			xmajorgrids=true,
			grid style=dashed,
			legend style={font=\tiny,legend columns=2},
			]
\addplot[
color=green,
mark=x,
very thin
]
coordinates {
(2.0, 840)
(2.2, 840)
(2.4, 840)
(2.6, 840)
(2.8, 840)
(3.0, 840)
(3.2, 840)
};	
\addlegendentry{O(2,8)}
\addplot[
color=violet,
mark=triangle,
very thin
]
coordinates {
(2.0, 1180)
(2.2, 1180)
(2.4, 1180)
(2.6, 1180)
(2.8, 1180)
(3.0, 1180)
(3.2, 1180)
};	
\addlegendentry{D-O(2,8)}
\addplot[
color=black,
mark=+,
very thin
]
coordinates {
(2.0, 11120)
(2.2, 11120)
(2.4, 11120)
(2.6, 11120)
(2.8, 11120)
(3.0, 11120)
(3.2, 11120)
};
\addlegendentry{ O(3,12)}
\addplot[
color=blue,
mark= halfcircle,
very thin
]
coordinates {
(2.0, 6228)
(2.2, 6279)
(2.4, 6360)
(2.6, 6404)
(2.8, 6496)
(3.0, 6578)
(3.2, 6657)
};	
\addlegendentry{D-O(3,12)-S}


\addplot[
color=purple,
mark=*,
very thin
]
coordinates {
(2.2, 520.40)
(2.4, 413.64)
(2.6, 284.33)
(2.8, 221.26)
(3.0, 145.98)
(3.2, 121.92)
};
\addlegendentry{PB$_0$}
\addplot[
color=magenta,
mark=square*,
]
coordinates {
(2.2, 1028.09)
(2.4, 797.60)
(2.6, 492.86)
(2.8, 383.88)
(3.0, 238.35)
(3.2, 206.42)
};
\addlegendentry{PB$_1$}
\addplot[
color=red,
mark=triangle*,
very thin
]
coordinates {
(2.0, 1524.33)
(2.2, 1092.50)
(2.4, 838.72)
(2.6, 509.43)
(2.8, 421.08 )
(3.0, 300.95)
(3.2, 238.19)
};	
\addlegendentry{PB$_2$}
\end{semilogyaxis}
\end{tikzpicture}
\caption{Average number of TEPs $n_{at}$ for LDPC (128,64) code under different schemes.}
\label{fer128_pb_average_tep}
\end{figure}	

%% file: sections/conclusions.tex
\section{Conclusions and Future Research}
\label{conclusions}

The proposed NMS-DIA-OSD architecture aims to achieve superior decoding performance, low latency, low complexity, high throughput, and independence from noise variance. To fully utilize the hidden information scattered in the decoding trajectories of the NMS decoder's failures, a bit-oriented deep learning model, termed DIA, is proposed to forge a new reliability metric after processing the bit-wise trajectory. Facilitated by the introduced DIA, the codeword-oriented conventional OSD is adapted to traverse a fixed sequence of test error patterns in a specific order, acquired before testing. The list size of the test error patterns can be further reduced either under the new proposed auxiliary criterion or by fine-tuned constraints on the blocks of test error patterns. Extensive simulations were conducted to compare with existing decoding schemes, verifying that the proposed architecture can effectively boost the decoding performance of short LDPC codes to near ML decoding, while retaining other targeted merits, thus emerging as a serious competitor to its counterparts. However, trials on long LDPC codes showed that it remains challenging for the proposed architecture to approach ML performance given limited computational resources. Another attempt on a traditional short BCH code demonstrated that no effective iterative message passing will prevent DIA from working properly, implying it fits the role of a post-processor for effective belief propagation (BP) variants.

Regarding the NMS decoding failures, the trajectories of bits jointly involved in each parity check of $\mathbf{H}$ can be treated as a multi-dimensional time series forecasting issue to forge a more reliable metric, thus warranting further investigation as an extension of DIA.

Not limited to the usage of CNNs, there are certainly more advanced neural network  structures available to implement DIA. Any off-the-shelf schemes for the time series forecasting task, such as temporal convolutional networks (TCNs) \cite{bai2018empirical}, are capable of constructing new bit reliability metrics and thus worth further exploration.

\section*{Availability of Data}
The parity check matrices of all related codes, as well as BP and ML decoding results, can be retrieved from the website \cite{helmling19} or \cite{Rosseel2022}. We are grateful for the efforts taken by the original authors for the code database maintenance and their generous sharing.

%% file: sections/ackowledgment.tex
\section*{Acknowledgments}
The authors would like to express their gratitude to Google Corporation for providing the excellent computing platforms of Colab and Kaggle online, which made it possible to debug, train, and test our models freely. Thanks are also extended to the anonymous reviewers for providing valuable feedback, contributing to the improvement of the quality of this paper.

%% file: main.bbl
\begin{thebibliography}{100}

\bibitem{gallager62}
Robert Gallager.
\newblock Low-density parity-check codes.
\newblock {\em IRE Trans. Inform. Theory}, 8(1):21--28, 1962.

\bibitem{mackay96}
David~JC MacKay and Radford~M Neal.
\newblock Near shannon limit performance of low density parity check codes.
\newblock {\em Electron. Lett.}, 32(18):1645, 1996.

\bibitem{zhao05}
Jianguang Zhao, Farhad Zarkeshvari, and Amir~H Banihashemi.
\newblock On implementation of min-sum algorithm and its modifications for decoding low-density parity-check (ldpc) codes.
\newblock {\em IEEE Trans. Commun}, 53(4):549--554, 2005.

\bibitem{jiang06}
Ming Jiang, Chunming Zhao, Li~Zhang, and Enyang Xu.
\newblock Adaptive offset min-sum algorithm for low-density parity check codes.
\newblock {\em IEEE Commun. Lett.}, 10(6):483--485, 2006.

\bibitem{yue2023efficient}
Chentao Yue, Vera Miloslavskaya, Mahyar Shirvanimoghaddam, Branka Vucetic, and Yonghui Li.
\newblock Efficient decoders for short block length codes in 6g urllc.
\newblock {\em IEEE Commun. Mag.}, 61(4):84--90, 2023.

\bibitem{Fossorier1995}
Marc~PC Fossorier and Shu Lin.
\newblock Soft-decision decoding of linear block codes based on ordered statistics.
\newblock {\em IEEE Trans. Inf. Theory}, 41(5):1379--1396, 1995.

\bibitem{wu2004new}
Xianren Wu, Hamid~R Sadjadpour, and Zhi Tian.
\newblock A new adaptive two-stage maximum-likelihood decoding algorithm for linear block codes.
\newblock In {\em 2004 IEEE Int. Conf. Commun. (IEEE Cat. No. 04CH37577)}, volume~2, pages 656--660. IEEE, 2004.

\bibitem{alnawayseh2012ordered}
Saif~EA Alnawayseh and Pavel Loskot.
\newblock Ordered statistics-based list decoding techniques for linear binary block codes.
\newblock {\em EURASIP J. Wirel. Commun. Netw.}, 2012:1--12, 2012.

\bibitem{yue2019segmentation}
Chentao Yue, Mahyar Shirvanimoghaddam, Yonghui Li, and Branka Vucetic.
\newblock Segmentation-discarding ordered-statistic decoding for linear block codes.
\newblock In {\em 2019 IEEE Global Commun. Conf. (GLOBECOM)}, pages 1--6. IEEE, 2019.

\bibitem{yue2021probability}
Chentao Yue, Mahyar Shirvanimoghaddam, Giyoon Park, Ok-Sun Park, Branka Vucetic, and Yonghui Li.
\newblock Probability-based ordered-statistics decoding for short block codes.
\newblock {\em IEEE Commun. Lett.}, 25(6):1791--1795, 2021.

\bibitem{choi2019fast}
Changryoul Choi and Jechang Jeong.
\newblock Fast and scalable soft decision decoding of linear block codes.
\newblock {\em IEEE Commun. Lett.}, 23(10):1753--1756, 2019.

\bibitem{cavarec2020learning}
Baptiste Cavarec, Hasan~Basri Celebi, Mats Bengtsson, and Mikael Skoglund.
\newblock A learning-based approach to address complexity-reliability tradeoff in os decoders.
\newblock In {\em 2020 54th Asilomar Conf. Signals, Syst., and Comput.}, pages 689--692. IEEE, 2020.

\bibitem{valembois2004box}
Antoine Valembois and Marc Fossorier.
\newblock Box and match techniques applied to soft-decision decoding.
\newblock {\em IEEE Trans. Inf. Theory}, 50(5):796--810, 2004.

\bibitem{yang2022reduced}
Lijia Yang, Wenhao Chen, and Li~Chen.
\newblock Reduced complexity ordered statistics decoding of linear block codes.
\newblock In {\em 2022 IEEE/CIC Int. Conf. Commun. in China (ICCC Workshops)}, pages 371--376. IEEE, 2022.

\bibitem{liang2022low}
Jifan Liang, Yiwen Wang, Suihua Cai, and Xiao Ma.
\newblock A low-complexity ordered statistic decoding of short block codes.
\newblock {\em IEEE Commun. Lett.}, 27(2):400--403, 2022.

\bibitem{yue2022linear}
Chentao Yue, Mahyar Shirvanimoghaddam, Giyoon Park, Ok-Sun Park, Branka Vucetic, and Yonghui Li.
\newblock Linear-equation ordered-statistics decoding.
\newblock {\em IEEE Trans. Commun.}, 70(11):7105--7123, 2022.

\bibitem{choi2020fast}
Changryoul Choi and Jechang Jeong.
\newblock Fast soft decision decoding of linear block codes using partial syndrome search.
\newblock In {\em 2020 IEEE Int. Symp. Inf. Theory (ISIT)}, pages 384--388. IEEE, 2020.

\bibitem{choi2021fast}
Changryoul Choi and Jechang Jeong.
\newblock Fast soft decision decoding algorithm for linear block codes using permuted generator matrices.
\newblock {\em IEEE Commun. Lett.}, 25(12):3775--3779, 2021.

\bibitem{yue2022ordered}
Chentao Yue, Mahyar Shirvanimoghaddam, Branka Vucetic, and Yonghui Li.
\newblock Ordered-statistics decoding with adaptive gaussian elimination reduction for short codes.
\newblock In {\em 2022 IEEE Globecom Workshops (GC Wkshps)}, pages 492--497. IEEE, 2022.

\bibitem{yue2021revisit}
Chentao Yue, Mahyar Shirvanimoghaddam, Branka Vucetic, and Yonghui Li.
\newblock A revisit to ordered statistics decoding: Distance distribution and decoding rules.
\newblock {\em IEEE Trans. Inf. Theory}, 67(7):4288--4337, 2021.

\bibitem{liang2024random}
Jifan Liang and Xiao Ma.
\newblock A random coding approach to performance analysis of the ordered statistic decoding with local constraints.
\newblock {\em arXiv:2401.16709}, 2024.

\bibitem{tataria20216g}
Harsh Tataria, Mansoor Shafi, Andreas~F Molisch, Mischa Dohler, Henrik Sj{\"o}land, and Fredrik Tufvesson.
\newblock 6g wireless systems: Vision, requirements, challenges, insights, and opportunities.
\newblock {\em Proc. of the IEEE}, 109(7):1166--1199, 2021.

\bibitem{shirvanimoghaddam2018short}
Mahyar Shirvanimoghaddam, Mohammad~Sadegh Mohammadi, Rana Abbas, Aleksandar Minja, Chentao Yue, Balazs Matuz, Guojun Han, Zihuai Lin, Wanchun Liu, Yonghui Li, et~al.
\newblock Short block-length codes for ultra-reliable low latency communications.
\newblock {\em IEEE Commun. Mag.}, 57(2):130--137, 2018.

\bibitem{Fossorier1999}
Marc~PC Fossorier, Miodrag Mihaljevic, and Hideki Imai.
\newblock Reduced complexity iterative decoding of low-density parity check codes based on belief propagation.
\newblock {\em IEEE Trans. Commun}, 47(5):673--680, 1999.

\bibitem{Fossorier2001}
Marc~PC Fossorier.
\newblock Iterative reliability-based decoding of low-density parity check codes.
\newblock {\em IEEE J. Sel. Areas Commun.}, 19(5):908--917, 2001.

\bibitem{jiang2007reliability}
Ming Jiang, Chunming Zhao, Enyang Xu, and Li~Zhang.
\newblock Reliability-based iterative decoding of ldpc codes using likelihood accumulation.
\newblock {\em IEEE Commun. Lett.}, 11(8):677--679, 2007.

\bibitem{baldi2016use}
Marco Baldi, Nicola Maturo, Enrico Paolini, and Franco Chiaraluce.
\newblock On the use of ordered statistics decoders for low-density parity-check codes in space telecommand links.
\newblock {\em EURASIP J. Wirel. Commun. Netw.}, 2016:1--15, 2016.

\bibitem{zhang2023efficient}
Weiyang Zhang, Chentao Yue, Yonghui Li, and Branka Vucetic.
\newblock Efficient near maximum-likelihood reliability-based decoding for short ldpc codes.
\newblock {\em arXiv:2306.00443}, 2023.

\bibitem{razzak18}
Muhammad~Imran Razzak, Saeeda Naz, and Ahmad Zaib.
\newblock Deep learning for medical image processing: Overview, challenges and the future.
\newblock {\em Classification in BioApps}, pages 323--350, 2018.

\bibitem{young18}
Tom Young, Devamanyu Hazarika, Soujanya Poria, and Erik Cambria.
\newblock Recent trends in deep learning based natural language processing.
\newblock {\em IEEE Comput. Intell. Mag.}, 13(3):55--75, 2018.

\bibitem{grigorescu20}
Sorin Grigorescu, Bogdan Trasnea, Tiberiu Cocias, and Gigel Macesanu.
\newblock A survey of deep learning techniques for autonomous driving.
\newblock {\em J. of Field Robot.}, 37(3):362--386, 2020.

\bibitem{nachmani16}
Eliya Nachmani, Yair Be'ery, and David Burshtein.
\newblock Learning to decode linear codes using deep learning.
\newblock In {\em 2016 54th Annu. Conf. Commun., Control, and Comput. (Allerton)}, pages 341--346. IEEE, 2016.

\bibitem{gruber17}
Tobias Gruber, Sebastian Cammerer, Jakob Hoydis, and Stephan ten Brink.
\newblock On deep learning-based channel decoding.
\newblock In {\em 2017 51st Annu. Conf. Inf. Sci. and Syst. (CISS)}, pages 1--6. IEEE, 2017.

\bibitem{nachmani18}
Eliya Nachmani, Elad Marciano, Loren Lugosch, Warren~J Gross, David Burshtein, and Yair Be’ery.
\newblock Deep learning methods for improved decoding of linear codes.
\newblock {\em IEEE J. Sel. Topics Signal Process.}, 12(1):119--131, 2018.

\bibitem{liang18}
Fei Liang, Cong Shen, and Feng Wu.
\newblock An iterative bp-cnn architecture for channel decoding.
\newblock {\em IEEE J. Sel. Topics Signal Process.}, 12(1):144--159, 2018.

\bibitem{lugosch18}
Loren Lugosch and Warren~J Gross.
\newblock Learning from the syndrome.
\newblock In {\em 2018 52nd Asilomar Conf. Signals, Syst., Comput.}, pages 594--598. IEEE, 2018.

\bibitem{lugosch18-1xVPf}
Loren~Peter Lugosch.
\newblock {\em Learning algorithms for error correction}.
\newblock McGill University (Canada), 2018.

\bibitem{wang20}
Qing Wang, Shunfu Wang, Haoyu Fang, Leian Chen, Luyong Chen, and Yuzhang Guo.
\newblock A model-driven deep learning method for normalized min-sum ldpc decoding.
\newblock In {\em 2020 Int. Conf. Commun. Workshops}, pages 1--6. IEEE, 2020.

\bibitem{helmling19}
Michael Helmling, Stefan Scholl, Florian Gensheimer, Tobias Dietz, Kira Kraft, Stefan Ruzika, and Norbert Wehn.
\newblock {D}atabase of {C}hannel {C}odes and {ML} {S}imulation {R}esults.
\newblock \url{www.uni-kl.de/channel-codes}, 2019.

\bibitem{nachmani22}
Eliya Nachmani and Yair Be’ery.
\newblock Neural decoding with optimization of node activations.
\newblock {\em IEEE Commun. Lett.}, 2022.

\bibitem{buchberger20}
Andreas Buchberger, Christian Häger, Henry~D Pfister, Laurent Schmalen, and Alexandre~Graell i~Amat.
\newblock Pruning and quantizing neural belief propagation decoders.
\newblock {\em IEEE J. Sel. Areas Commun.}, 39(7):1957--1966, 2020.

\bibitem{buchberger21}
Andreas Buchberger, Christian Häger, Henry~D Pfister, Laurent Schmalen, and Alexandre~Graell i~Amat.
\newblock Learned decimation for neural belief propagation decoders.
\newblock In {\em 2021-2021 Int. Conf. Acoust., Speech and Signal Process. (ICASSP)}, pages 8273--8277. IEEE, 2021.

\bibitem{Rosseel2022}
Joachim Rosseel, Valérian Mannoni, Inbar Fijalkow, and Valentin Savin.
\newblock Decoding short ldpc codes via bp-rnn diversity and reliability-based post-processing.
\newblock {\em IEEE Trans. Commun.}, 70(12):7830--7842, 2022.

\bibitem{chollet2015keras}
Fran\c{c}ois Chollet et~al.
\newblock Keras.
\newblock \url{https://keras.io}, 2015.

\bibitem{kingma14}
Diederik~P Kingma and Jimmy Ba.
\newblock Adam: A method for stochastic optimization.
\newblock {\em arXiv:1412.6980}, 2014.

\bibitem{bai2018empirical}
Shaojie Bai, J~Zico Kolter, and Vladlen Koltun.
\newblock An empirical evaluation of generic convolutional and recurrent networks for sequence modeling.
\newblock {\em arXiv:1803.01271}, 2018.

\end{thebibliography}
